\documentclass[10pt,a4paper]{article}
\usepackage[top=2cm, bottom=2cm, left=2cm, right=2cm]{geometry}
\usepackage[super,sort,compress]{cite}
\usepackage{multirow}%
\usepackage{amsmath,amssymb,amsfonts}%
\usepackage{amsthm}%
\usepackage{mathrsfs}%
\usepackage[title]{appendix}%
\usepackage{xcolor}%
\usepackage{textcomp}%
\usepackage{manyfoot}%
\usepackage{booktabs}%
\usepackage{algorithm}%
\usepackage{algorithmicx}%
\usepackage{algpseudocode}%
\usepackage{listings}%

\usepackage{xspace} 
\usepackage{graphicx}
\usepackage{sidecap}
\usepackage{caption}
\usepackage{subcaption}

\usepackage{color,soul}
\usepackage{xcolor}
\usepackage{authblk}

\usepackage[]{mdframed}
\DeclareCaptionFormat{custom}
{%
    \textbf{#1#2}{\small #3}
}
\DeclareCaptionLabelSeparator{custom}{\,$\mid$\,}
\title{Excitonic signatures of ferroelectric order in parallel-stacked MoS$_2$}
\date{}

\begin{document}
\author[1]{Swarup Deb*}
\author[1]{Johannes Krause}
\author[4]{Paulo E. {Faria~Junior}}
\author[1]{Michael Andreas Kempf}
\author[1]{Rico Schwartz}
\author[2]{Kenji Watanabe}
\author[3]{Takashi Taniguchi}

\author[4]{Jaroslav Fabian}
\author[1]{Tobias Korn**}

\affil[1]{\small Institute of Physics, University of Rostock, Albert-Einstein-Str. 23, Rostock 18059, Germany}	
\affil[2]{Research Center for Electronic and Optical Materials, NIMS, 1-1 Namiki, Tsukuba 305-0044, Japan}
\affil[3]{Research Center for Materials Nanoarchitectonics, NIMS, 1-1 Namiki, Tsukuba 305-0044, Japan}
\affil[4]{Institute for Theoretical Physics, University of Regensburg, 93040 Regensburg, Germany}
\affil[*]{swarupdeb2580@gmail.com}
\affil[**]{tobias.korn@uni-rostock.de}

\maketitle
\begin{abstract}
Interfacial ferroelectricity, prevalent in various parallel-stacked layered materials, allows switching of out-of-plane ferroelectric order by in-plane sliding of adjacent layers. Its resilience against doping potentially enables next-generation storage and logic devices. However, studies have been limited to indirect sensing or visualization of ferroelectricity. For transition metal dichalcogenides, there is little knowledge about the influence of ferroelectric order on their intrinsic valley and excitonic properties. Here, we report direct probing of ferroelectricity in few-layer 3R-MoS$_2$ using reflectance contrast spectroscopy. Contrary to a simple electrostatic perception, layer-hybridized excitons with out-of-plane electric dipole moment remain decoupled from ferroelectric ordering, while intralayer excitons with in-plane dipole orientation are sensitive to it. Ab initio calculations identify stacking-specific interlayer hybridization leading to this asymmetric response. Exploiting this sensitivity, we demonstrate optical readout and control of multi-state polarization with hysteretic switching in a field-effect device. Time-resolved Kerr ellipticity reveals direct correspondence between spin-valley dynamics and stacking order. 
\end{abstract}

\subsection*{Introduction}
The basic building block for any interfacial ferroelectrics is two layers of certain van der Waals materials aligned parallel to each other~\cite{li2017binary,woods2021charge, yasuda2021stacking, vizner2021interfacial, wang2022interfacial, weston2022interfacial,deb2022cumulative, rogee2022ferroelectricity, garcia2023mixed,yang2023atypical,atri2023spontaneous}. In the case of transition metal dichalcogenides (TMDs) like MoS$_2$, layer-asymmetric atomic registry\cite{cao2022interlayer} along the out-of-plane direction leads to interlayer hybridization between the valence band of one layer and the conduction band of the other but not vice versa~\cite{ferreira2021weak}. Such a preferential coupling induces unidirectional charge transfer, leading to the spontaneous emergence of an electric dipole bound at the interface. A bilayer unit of parallelly-stacked (rhombohedral or so-called 3R)-TMDs thus possesses two equivalent yet opposite ferroelectric orders viz. MX and XM, marking the stacking of metal (M) and chalcogen (X) atoms at the eclipsed sites (Fig. \ref{fig:1}).  

Besides hosting ferroelectricity, parallel stacking of TMDs preserves the spin-valley locking, regarded predominantly as a property of monolayers~\cite{Xiao2012},  even in the multilayer limit due to broken inversion and mirror symmetries\cite{suzuki2014valley,sung2020broken,xu2022towards,dong2023giant}. Therefore, the 3R-polymorph of TMDs provide means for the bottom-up construction of three-dimensional spin-valleytronic devices on a ferroelectric platform. A visionary goal would be to engineer the optical response in TMDs by exploiting the ferroelectricity-induced interaction, which can be highly localized, non-volatile, and reconfigurable. Despite a tremendous interest in this emerging phenomenon \cite{fox2023stacking, naumis2023mechanical}, so far the field has been  limited to indirect sensing or visualizing the ferroelectricity employing surface-sensitive probes, such as atomic force microscopy~\cite{woods2021charge, vizner2021interfacial,  weston2022interfacial,deb2022cumulative,  atri2023spontaneous}, scanning electron microscopy~\cite{ko2023operando}, or sensing-layer-based approaches~\cite{woods2021charge,wang2022interfacial} and, therefore, could only provide limited physical insight.  In the case of TMDs, this leaves a void of in-depth knowledge about the influence of ferroelectric order on their intrinsic valley and excitonic properties. 
To fill this gap, we exploit hyperspectral reflectivity imaging to discern the correspondence between ferroelectric stacking and optical response in these structures. We demonstrate ferroelectricity-induced control of spin-valley dynamics by systematically studying the ultrafast response of excitons using transient reflection and Kerr ellipticity. As-grown samples are chosen over artificially stacked parallel structures, which provides a moir\'e-free crystal, enabling us to explore excitonic phenomena of purely ferroelectric origin and to employ far-field optical spectroscopy on mesoscopic domains with well-defined stacking. Clear signatures of stacking-dependent K-valley excitonic response and a striking contrast in valley relaxation dynamics are observed. This presents the prospect of using spin and valley as useful degrees of freedom in a robust multilayer platform. It is worth mentioning that recent experimental efforts \cite{tyulnev2024valleytronics} also underscore the possibility of utilizing multilayer 2H-TMDs for valleytronic applications despite their centro-symmetric structure\cite{paradisanos2020controlling}.

\subsection*{Results}
\subsubsection*{Ferroelectric domains in 3R single crystals}

We begin by probing the electric surface potential of deterministically stamped 3R-MoS$_2$ flakes on non-polar hexagonal boron nitride (\textit{h}BN) exfoliated on Si/SiO$_2$(-285\,nm) substrate (SI.S1). An atomic force microscope operated in side-band Kelvin probe mode (KPFM) (SI.S2) is used to map the surface potential, V$_{\text{KP}}$. Fig. \ref{fig:1}a shows an optical microscopy image of a representative flake (Sample\,1) composed of five and seven layers of MoS$_2$. In a multilayer flake, each of the interfaces may have either XM (polarization pointing upward) or MX (downward) stacking order creating a potential ladder either going upwards or downwards, respectively\cite{ferreira2021weak, deb2022cumulative}. The difference in the total number of upward- and downward-pointing polarized interfaces decides the cumulative polarization state, hence the surface potential. Therefore, different ferroelectric domains in a given thickness would be visible in  a \textit{V}$_{KP}$ map. Based on this understanding, the existence of two ferroelectric domains, viz. domain I and II in the 5L region of the sample, becomes evident from Fig. \ref{fig:1}b. The measured surface potential variation of $\sim$127\,mV between these domains (SI Fig. S1) corresponds to the sum of effectively two ferroelectric interfaces, consistent with  earlier results\cite{deb2022cumulative, cao2024polarization}.

\begin{mdframed}
Box 1: Supplementary \textbf{Figure S2}	
\centering
\includegraphics[scale = 0.4]{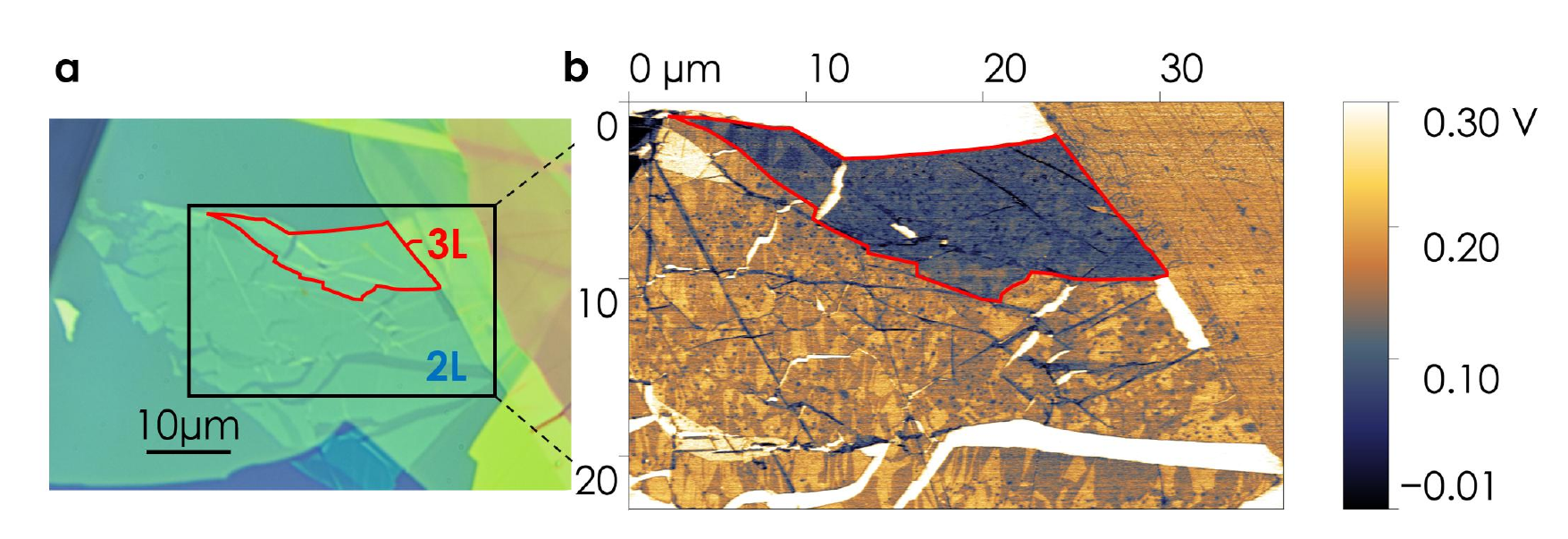}
\renewcommand{\thefigure}{{\small }}
\captionof{figure}{\textbf{Inducing ferroelectric domains in naturally grown 3R-MoS$_2$.} \textbf{a.} Optical micrograph of a bi- and trilayer 3R-MoS$_2$ flake stamped on Si/SiO$_2$/\textit{h}BN with deliberate shear perturbation (horizontal force) during the transfer from PDMS. \textbf{b.} Surface potential map within the area enclosed by the black rectangle in a. Densely packed ferroelectric domains can be identified by the difference in contrast. (partial copy of SI Fig.S2)}
\end{mdframed}

The appearance of different domains in a single crystal flake can be attributed either to inherent stacking faults in the bulk 3R crystal or to the sliding by slip-avalanche of constituting layers from XM to MX stacking order or vice versa. Since the energies of all possible stacking configurations are similar (SI Fig. S19), the latter occurs as a result of inadvertent shear strain during exfoliation~\cite{liang2023shear} or deterministic stamping. To further illustrate the mechanism, we fabricated a sample by intentionally applying an in-plane perturbation, achieved by laterally tapping the substrate holder during the dry transfer of 3R-MoS$_2$ from PDMS to the substrate. This horizontal force introduces a shear stress between the layers that remain anchored to PDMS and the layers attached to the substrate. This deliberate horizontal force, which is significantly stronger than the ubiquitously present shear forces during standard operations, generates smaller and densely packed ferroelectric domains, as identified by KPFM (SI Fig. S2 and the box 1). The ferroelectric domains, in this case, are of similar dimension as in the artificially stacked moir\'e interfacial ferroelectrics. However, achieving precise control over domain formation and their structural organization remains a challenge.

\subsubsection*{Correspondence of surface potential and exciton fine structure}

\begin{figure*}
	\setcounter{figure}{0}
	\centering
	\includegraphics[scale=0.65]{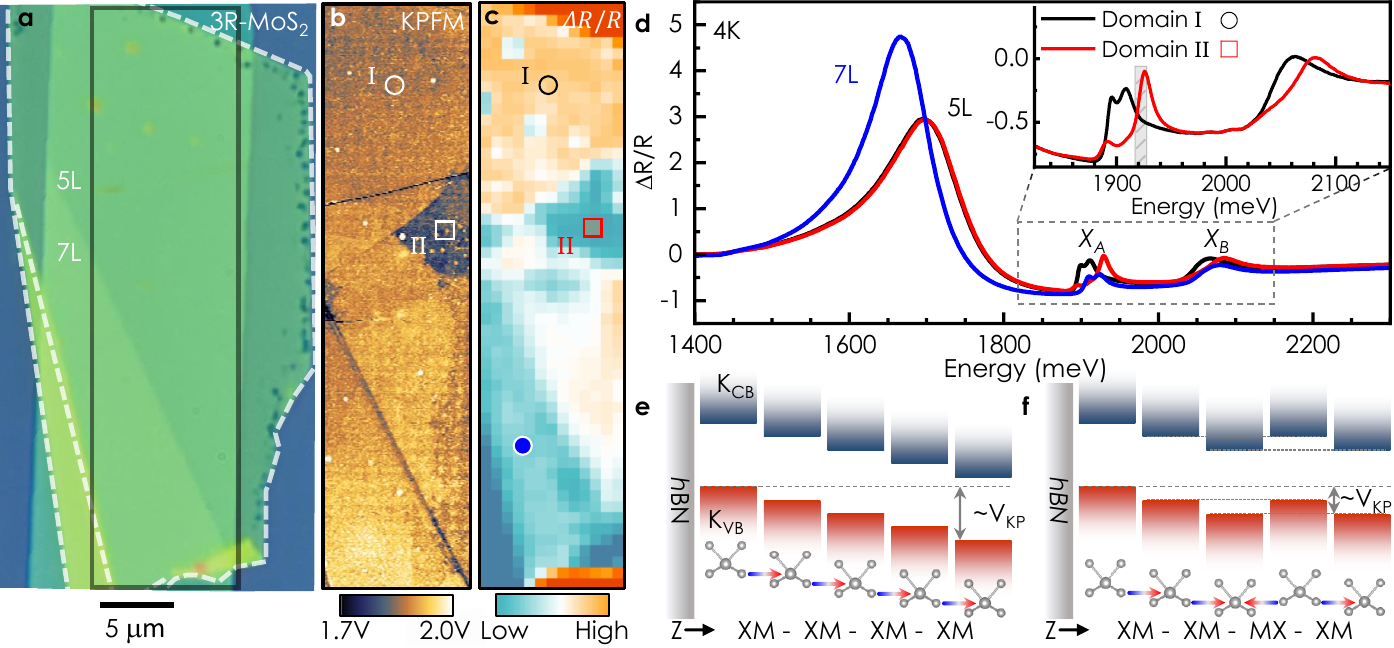}
	\caption{\textbf{Reflection contrast imaging of ferroelectric domains in  few-layer 3R-MoS$_2$.} \textbf{a.} Optical micrograph of a 3R-MoS$_2$ flake on Si/SiO$_2$/\textit{h}BN. Topographical steps and edges of the bottom \textit{h}BN have been marked by white dotted lines. \textbf{b.} Surface potential map within the area enclosed by the black rectangle in a. Two domains, marked by {\large{$\circ$}} and $\Box$, can be identified by the difference in contrast. \textbf{c.} Integrated intensity map of $\Delta R/R$ at the $X_A$ spectral region, i.e., from 1906 to 1918\,meV (gray bar in inset of panel d). \textbf{d.} Low-temperature reflectance contrast spectra from various spatial locations marked by symbols of corresponding color in c (black - 5L domain I, red - 5L domain II, and blue - 7L). Inset shows high-resolution spectra collected from domain I and II. \textbf{e-f.} Two crystal configurations of 5L 3R-MoS$_2$ overlayed on layer-projected K$_{\text{VB}}$ and K$_{\text{CB}}$ band-edges.  To first order, the band edge variation along Z and the degeneracies, highlighted for the XM-XM-MX-XM stacking by dotted lines,  result from the ferroelectricity-induced electrostatic considerations (SI Fig. S12-13 for other configurations). Arrows represent polarization vectors at the interfaces.}
	\label{fig:1}
\end{figure*}

To check for the ferroelectric-field-induced changes in optical response, we map the exciton transition energies over the entire flake (i.e. Sample 1) using confocal differential reflectance spectroscopy (SI.S3). Typical reflectance contrast (RC) spectra, $\Delta R/R = (R_{Ref} - R_{Flake})/R_{Flake}$\cite{frisenda2017micro} obtained at different spatial locations of the flake at liquid helium temperature are presented in Fig. \ref{fig:1}d. Here $R_{Flake}$ is the reflection from MoS$_2$ and $R_{Ref}$ is the same from \textit{h}BN (SI Fig. S3). In the energy range of interest, the reflectance contrast of the few-layer MoS$_2$ is dominated by three excitonic features, viz. the momentum-indirect transition at $\sim$1700 meV and the well-known A and B excitons at \textit{X}$_A \sim$1900 and \textit{X}$_B \sim$2050 meV \cite{wang2018colloquium,li2022stacking, ullah2021selective}. Notably, the lowest energy feature despite being a momentum-indirect transition has a prominent appearance. This can be attributed to the favorable thin-film interference condition in the Fabry-P\'erot cavity formed by Si/SiO$_2$/\textit{h}BN/MoS$_2$ in the given energy range (SI Fig. S4 - S5 for additional examples and further discussion). We note that RC spectra, $\Delta R/R$ can also be calculated using $ (R_{Ref} - R_{Flake})/R_{Ref}$\cite{mcintyre1971differential}. We found that both representations, $\Delta R/R_{Ref}$ and $\Delta R/R_{Flake}$ share similar characteristic viz. the presence of three distinct features and their energetic positions (SI Fig. S4, S5, S6, S15). However, the choice of definition modifies the line shape of each transition and therefore, the overall appearance of the spectra. Evidently, the lowest energy peak appears vividly in $\Delta R/R_{Flake}$ (SI Fig. S4 - S5, S10). Therefore, in the following we refer to $\Delta R/R_{Flake}$ as RC spectra unless mentioned otherwise.

\begin{SCfigure}
	\centering
	\includegraphics[scale=0.5]{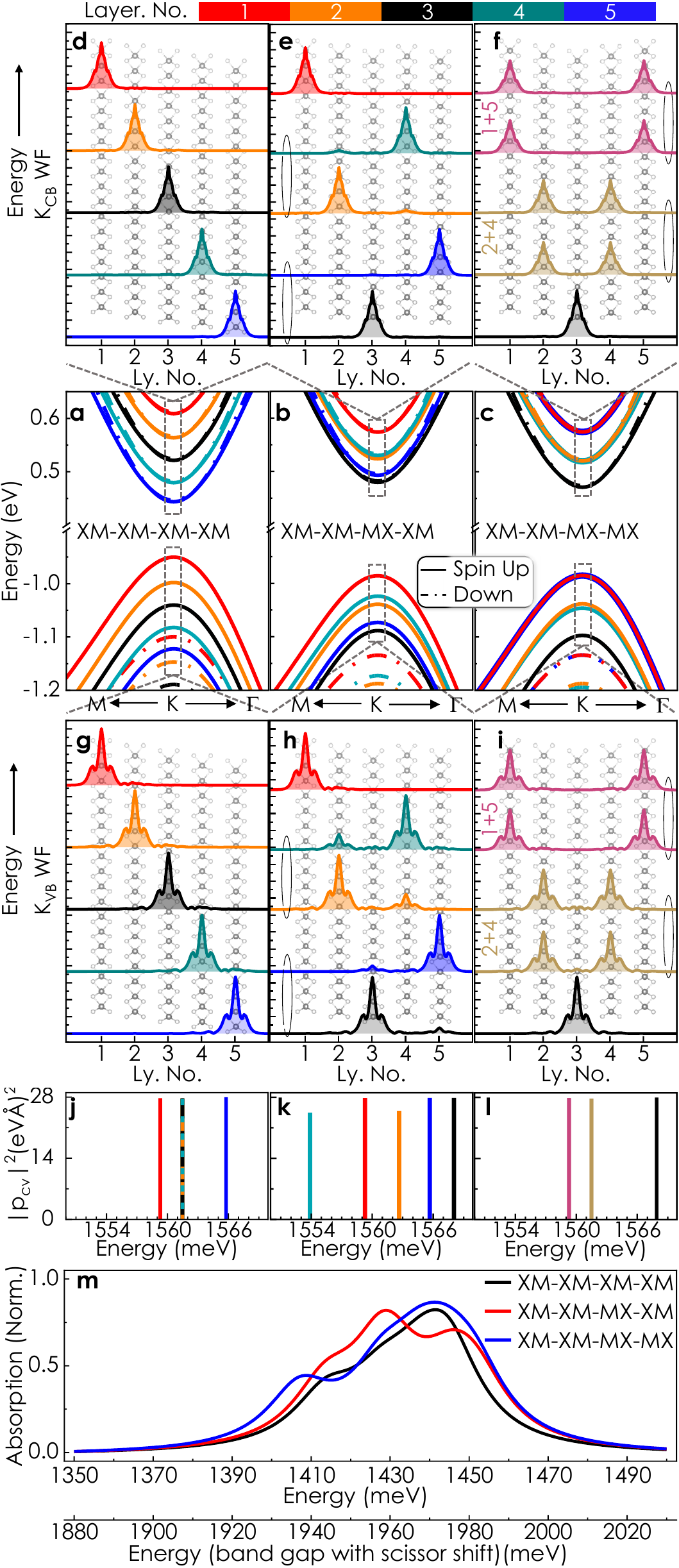}
	\caption{\textbf{Ab initio calculation of stacking-dependent band structure, wavefunctions and optical response.} \textbf{a-c.} Band structure  for different stacking orders. Spin orientation ($\parallel$\textit{Z}) of the individual bands is indicated by solid (spin up) and dashed (spin down) lines, respectively. The bands are colored corresponding to the dominant contribution of specific layers. The stacking of the band extrema at K follows the underlying ferroelectric potential. Due to mirror symmetry for the XM-XM-MX-MX case,  some bands are (nearly) degenerate and cannot be attributed to an individual layer (indicated by different colors in panel f, i).  \textbf{d-i.} Absolute value of the lowest-energy spin-up conduction band (d-f) and highest-energy valence band (g-i) wavefunctions sorted according to their energies superimposed over the crystal structure. Colors indicate the dominant layer(s) for each wavefunction. Ellipses are used to indicate (near-)degenerate states. The wave functions are layer-localized for the fully polarized case (d, g). For XM-XM-MX-MX (f, i), the lowest-energy wave function is localized in the center layer, while the other wave functions have equal weight in the mirror-symmetric layers. \textbf{j-l.} Oscillator strength for transitions from all spin-up K$_{\text{VB}}$ to all  K$_{\text{CB}}$ with $s^+$ polarization. The magnitude of these intralayer transitions is in good agreement with values for pristine monolayer MoS$_2$\cite{Wozniak2020PRB,FariaJunior2022NJP}. \textbf{m.}  Calculated absorption spectra in the A excitonic region based on intralayer dipole matrix elements and layer-specific exciton binding energy,  including a phenomenological Lorentzian line width of 20~meV.}
	\label{fig:2}
\end{SCfigure}

We proceed by constructing RC maps corresponding to different transitions by integrating the intensity  over selected spectral ranges. The map of the momentum-indirect transition brings out a clear contrast between the five- and seven-layer regions (SI Fig. S7 - S8) due to the red-shift in its spectral position with increasing thickness. On the other hand, the $X_A$ intensity map exhibits signatures of two distinct domains  within the topographically smooth 5L area, similar to the KPFM map. The contrast in the intensity map (Fig. \ref{fig:1}c) is due to the appearance of higher- and lower-energy sub-features in the spectral region from 1880 to 1930 meV. The energy positions and relative oscillator strengths of these fine features distinguishably mark the ferroelectric stacking configuration, as revealed by the high-resolution spectra shown in Fig. \ref{fig:1}d-inset. In contrast to previous studies on artificial R-stacked TMDs~\cite{Zhao2023}, these subfeatures are not related to nanoscale reconstruction of moir\'e domains (as confirmed by the complementary KPFM measurements). A similar spatial map is obtained from room temperature spectroscopy measurements (SI Fig. S9); however, due to thermal broadening, spectral features can not be fully resolved. Correspondence between RC maps and KPFM images has been observed in several samples of different thicknesses, see Fig. S10 for example. Upon closer inspection, it becomes evident that $X_B$ exhibits a similar trait (SI Fig. S7 - S8). To improve the signal-to-noise ratio of the RC spectra around the $X_B$ transition, we exploit the cavity effect by carefully choosing the bottom \textit{h}BN of our stack. SI Fig. S11 shows an RC spectrum of a trilayer-MoS$_2$ flake on a 95\,nm-\textit{h}BN/285\,nm-SiO$_2$/Si substrate, where the splitting in the $X_B$ becomes apparent.  In a nutshell, momentum-direct excitons at the K points in the Brillouin zone reveal the stacking order, while the spectral feature related to the momentum-indirect transition remains insensitive to it. Based on the trend of red-shift \cite{splendiani2010emerging} with increasing layer number, as shown in Fig. S4 and previous studied on similar systems \cite{ullah2021selective, khestanova2023robustness}, one can tentatively assign the low-energy peak either to the $\Gamma$-Q or to K-Q hybrid-excitonic \cite{tagarelli2023electrical} transition.

To explain the asymmetric response of different exciton species to ferroelectric ordering, we consider the band structure of 3R-MoS$_2$. It has been shown that in bilayer 3R-TMDs, states at the $\Gamma$ point of the valence band (\text{VB}) or Q point of conduction band (\text{CB}) are fully delocalized due to strong interlayer hybridization~\cite{deb2022cumulative, ferreira2021weak, sung2020broken}. In contrast, the wavefunctions at the K points are almost entirely layer-localized. The ferroelectricity-induced built-in potential manifests as a local electrostatic perturbation and introduces a shift between the layer-projected energy levels at K points, leading to a type-II band alignment at the interfaces~\cite{sung2020broken}, with the energies decreasing from layer X to layer M. By the same token, in a multilayer sample, the spatial profile of K-point band extrema along the out-of-plane direction follows the underlying ferroelectric potential and is, therefore, susceptible to the stacking order, as depicted schematically in Fig. \ref{fig:1}e-f and SI Fig. S12 - S13.

Neglecting interlayer coupling through hybridization, a built-in electric field should shift the conduction and valence bands by the same amount, leaving the transition energies unchanged. Clearly, this alone is insufficient for interpreting the experimentally observed emerging multiplicity of transition energies. Therefore, to this end, we inspect the stacking-dependent strength of K states delocalization based on our band structure calculations depicted in Fig. \ref{fig:2}a-c. Figure \ref{fig:2}d-f, g-i presents the five lowest energy conduction-band and highest-energy valence-band spin-up  electron wavefunctions at the K point, sorted according to their energy eigenvalues (for the following sections, we have dropped the term `K point' while referring to bands and wavefunctions unless stated otherwise). For the fully polarized, XM-XM-XM-XM case (Fig. \ref{fig:2} a,d,g), the individual wave functions both at the valence and conduction band remain layer-localized with their energy levels corresponding to the ferroelectric potential ascending with each additional layer. However, for other variants, interlayer hybridization is evident (e.g., Fig \ref{fig:2}b,e,h and c,f,i). This dependence of hybridization on ferroelectric ordering is the underlying mechanism for the observed contrast in optical spectra. Before proceeding further, we note that a five-layer 3R-MoS$_2$ crystal can have sixteen ($2^{N-1}$, $N=5$) stacking configurations, out of which KPFM can distinguish among five combinations based on their surface potential (SI Fig. S12). At this stage, therefore, instead of attempting a one-to-one quantitative description of the stacking-dependent excitonic response, we focus on the qualitative generic trend with three specific cases as examples (all the studied systems are shown in the SI Fig. S20 - S27).

In the fully co-polarized XM-XM-XM-XM stacking, the potential varies monotonically from layer to layer\cite{deb2022cumulative}. Thus, the three central layers experience a similar ferroelectric field - preserving the energy degeneracy of their intralayer excitonic transitions, appearing as a single vertical line in the dipole matrix element plot (Fig. \ref{fig:2}j). However, the top and bottom layers show energetically distinct transitions due to broken translational symmetry at the surfaces. The scenario changes in partially polarized stacking due to selective interlayer coupling between specific layers. The mechanism described below is similar to the quantum correction that leads to level anti-crossing by lifting the energy degeneracy of coupled quantum systems. For instance, in the XM-XM-MX-XM configuration, electrostatic degeneracy (cf. Fig. \ref{fig:1}f) leads to layer-specific resonant hybridization \cite{PhysRevB.42.1841} of K$_{\text{VB}}$ states, which is particularly strong between the second and fourth layers. Pronounced hybridization also occurs in the XM-XM-MX-MX case between the mirror-symmetric layers, namely the first and fifth, as well as the second and fourth. In contrast to the valence band states, the K$_{\text{CB}}$ states largely retain their layer-localized nature in almost all cases except mirror-symmetric stacking orders. This layer- and band-specific strength of interlayer hybridization leads to unequal shifts of layer-projected conduction and valence bands, lifting the intralayer transition degeneracy even for the center layers and increasing the maximum spread of energies, clearly visible in Fig. \ref{fig:2}j-l. These findings are of a general nature for any parallel-stacked multilayer TMDs. This stacking-specific resonant hybridization is not relevant for the $\Gamma$ and Q states, which have a significant degree of hybridization irrespective of stacking.

In addition to the corrections to the electronic band structure arising from hybridization, changes in the exciton binding energy due to asymmetric dielectric surroundings also play a role in determining the energetic position of individual transitions\cite{ZiliangPRX}. In order to provide a qualitative comparison to the experimental reflectance contrast spectra, we calculate the absorption spectra based on the intralayer dipole matrix elements and exciton binding energy (SI.S5 - S6),  considering a phenomenological Lorentzian broadening with a full-width-half-maximum of 20\,meV. These results are depicted in Fig. \ref{fig:2}m. Evidently, the general shape with multiple sub-features, which strongly depend on the individual stacking, closely mimics the experimental observations. However, it is important to note that DFT typically underestimates the band gap of semiconductors due to the unaccounted derivative discontinuity of the typical exchange-correlation functionals\cite{GodbyPRB} and, therefore, the calculated absorption spectra are red-shifted by about 480\,meV in comparison with the experimental results. By applying a rigid energy shift to the band gap, estimated from GW calculations, one can achieve transition energies that are more consistent with experimental data. For MoS$_2$, this estimate yields a band gap increase of about 530\,meV\cite{Kim2021PRB}, which nearly matches the experimentally observed energies.

These observations provide an alternative experimental approach to identify local changes in ferroelectric ordering in 3R-TMD flakes without resorting to surface-sensitive\cite{wang2022interfacial,weston2022interfacial,deb2022cumulative,zhang2023visualizing}  measurements, which become a challenge for heterostructures. In contrast to other optical measurement schemes relying on a sensing layer\cite{fraunie2023electron}, the TMD itself reveals its ferroelectric order in optical spectra. This allows for identifying local domains, even in `buried' layers, as required for integrating more complex, functional heterostructures based on ferroelectric TMDs.

\subsubsection*{Non-volatile electrical control and optical readout of multi-state polarization}

\begin{SCfigure}
	\centering
	\includegraphics[scale=0.55]{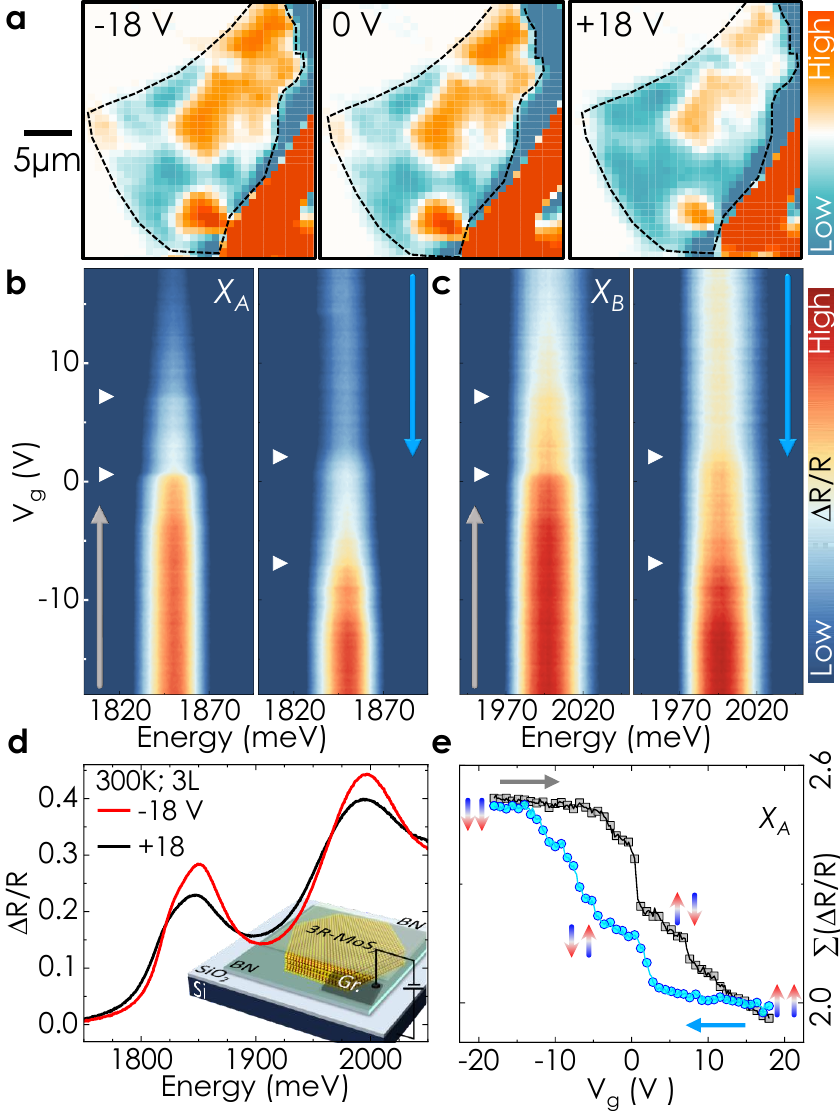}
	\caption{\textbf{Hysteretic reflectance contrast as a function of gate voltage.} \textbf{a.}  Integrated amplitude false color map of the numerically calculated first derivative of RC spectra at different gate voltages to visualize domain switching. All measurements were performed at room temperature. Evidently, the out-of-plane electric field drives the blue-colored region  to grow in area coverage with increasing gate voltage at the expense of the orange-colored region. \textbf{b-c.} False color map of A and B exciton intensity as a function of gate voltage during forward and backward sweep. They share the same y-axis, given on the left. \textbf{d.} Two representative room temperature reflectance contrast spectra at opposite gate voltages.  Inset- schematic illustration of the field effect device. \textbf{e.} Integrated intensity profile ($\equiv$ vertical line cuts from panel b) at 1852$\pm$2\,meV. The vertical arrows represent the ferroelectric ordering of each interface.}
	\label{fig:3}
\end{SCfigure}

Our findings, therefore, present an unprecedented opportunity of exploiting ferroelectric fields to gain non-volatile control over excitonic transitions in a field effect transistor architecture- a crucial step forward for applications. We prepare devices on a conducting Si/SiO$_2$(-90\,nm) substrate, which acts as the gate electrode, and use few-layer graphite flakes to make electrical contact with 3R-MoS$_2$ (viz. Sample 4). We use fully \textit{h}BN-encapsulated structures for these measurements (therefore, $R_{Flake}$ is the reflection from \textit{h}BN/MoS$_2$/\textit{h}BN heterostructure and $R_{Ref}$ is the same from \textit{h}BN/\textit{h}BN). Here, we present results obtained on a trilayer sample at room temperature. First, we map the ferroelectric domains at a few different gate voltages. Two different domains are visible in the false color map as blue and orange regions. Application of positive gate voltage results in an expansion of the blue domains at the cost of the orange domains. An opposite change is observed for negative gate voltage. Next, we measure the reflection contrast close to a domain boundary as a function of gate voltage. The $X_A$ and $X_B$ transitions becomes weaker as the gate voltage is swept forward from -18 V to 18 V, corresponding to increasing electron doping. This reduction is due to a combined effect of Pauli blocking and reduced oscillator strength with state filling - typical of n-type TMDs\cite{newaz2013electrical, wang2017valley, li2021refractive}. Notably, the reduction of intensity from the highest to lowest occurs through two distinct steps during the sweeping process (Fig. \ref{fig:3}e). Furthermore, a prominent hysteretic behavior in the intensity of $X_A$ and $X_B$ transitions is observed  as the gate voltage decreases during a backward sweep, for which the intensity rises again and saturates  (Fig. \ref{fig:3}b-e and SI Fig. S16). We also examine the standard deviation, which correlates with the full-width half maximum of $X_A$ and $X_B$. Evidently, the resulting plots exhibit the same features (SI Fig. S17).

The step-like features indicate discrete changes in optical response due to ferroelectric switching through a mechanism known as slidetronics, where the out-of-plane electric field induces an in-plane sliding of adjacent layers\cite{vizner2021interfacial, meng2022sliding, ko2023operando,  van2024engineering, sui2024atomic}. At high electric fields, all the interfaces align to achieve an energetically favorable configuration, viz. MX-MX or XM-XM, as indicated in Fig. \ref{fig:3}e. Therefore, the hysteretic occurrence of the discrete switching events in response to the external field directly results from a finite coercive field - a hallmark of ferroelectric materials. The two distinctive steps observed for the trilayer sample reveal that switching between the multiple ferroelectric states of a multilayer is feasible and controllable via external voltages - a pivotal insight for scalability.

Our study yields an average coercive field of 0.03\,-\,0.035\,Vnm$^{-1}$ for domain wall motion. Moreover, the hysteretic behavior of the monotonous intensity variation with gate voltage indicates a stacking-dependent change in average carrier concentration, observed earlier in transport measurements\cite{weston2022interfacial, lv2024multiresistance}. We note that a complete ferroelectric switching is not achieved throughout the whole crystal within the experimentally accessible gate voltage. This can be attributed to the  increasing domain boundary deformation energy\cite{enaldiev2022scalable} and domain wall pinning at localized stacking faults as well as surface contaminants.\cite{ko2023operando}

\subsubsection*{Domain-resolved exciton population and spin-valley dynamics}

\begin{figure*}[h]
	\centering
	\includegraphics[scale=0.55]{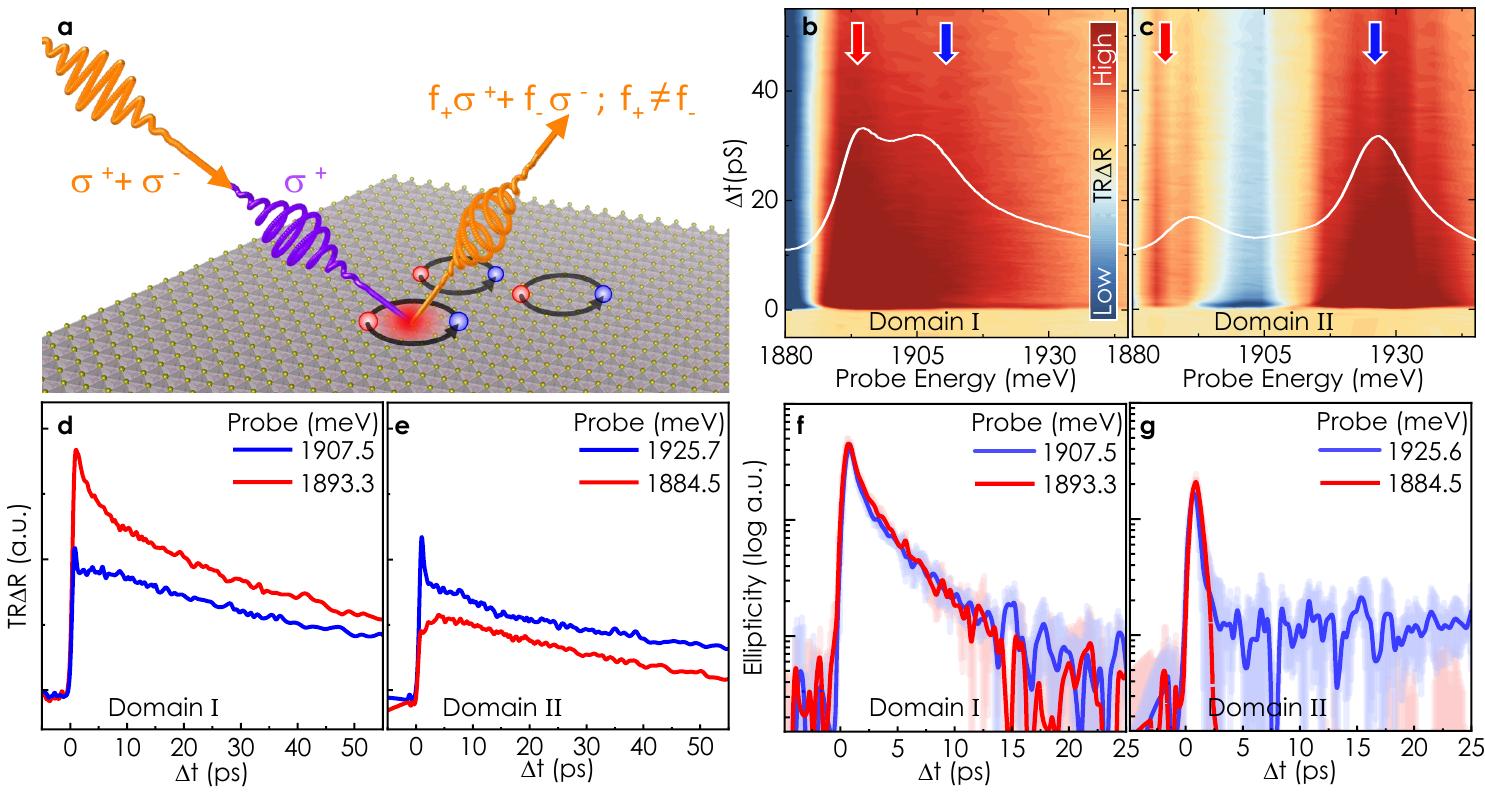}
	\caption{\textbf{Exciton population and spin dynamics.} \textbf{a.} Schematic of pump-probe Kerr measurement. A blue detuned (to $X_A$) circularly polarized light pulse for the pump and linearly polarized pulse for the probe were used. Upon reflection, the linearly polarized pulses turn elliptically polarized. By projecting the reflected elliptic pulse on a quarter waveplate and a Wollaston prism, we can separate out the relative strength of right and left circular components. The total probe intensity change is equivalent to net exciton density, and the intensity difference between left and right circular components gives the net spin-valley polarization. All the pump-probe measurements were done at 4\,K. \textbf{b-c.} False color map of transient reflection as a function of probe energy from domain I and II. White contours (without y-scale) are copied from Fig.1d-inset for comparison. \textbf{d-e.} Transient reflection at selected energies, marked by colored arrows in b and c. The opposite trend in the amplitude of the TR$\Delta$R signal at the lower and higher energy states from the two domains is a direct consequence of opposite relative strength of these sub-features in the steady-state spectroscopic signal. \textbf{f-g.} Traces of transient ellipticity at resonant energies in respective domains. The spin-valley lifetime in domain I is on the order of 10-15~ps, while in domain II the signal decays down to the detection noise-floor on sub-ps timescales.}
	\label{fig:4}
\end{figure*}

Besides gaining control over the excitonic oscillator strength, a meticulous understanding of excitonic lifetime is essential in the pursuit of realizing exciton-based circuits. To this end, we focus on probing the stacking-order-dominated exciton dynamics and spin relaxation using two-color pump-probe Kerr microscopy. Figure \ref{fig:4}a illustrates the measurement scheme. We use a circularly polarized pump pulse, slightly blue detuned to $X_A$, to induce a spin polarization of chosen helicity. The photo-generated excitons lead to photo-bleaching of a given valley, resulting in helicity-dependent reflectivity. Therefore, the transient exciton population and its spin-valley polarization can be probed by recording the intensity and induced ellipticity of a time-delayed linearly polarized probe pulse after reflection\cite{Kempf2021, kempf2023rapid}. A series of simultaneous transient reflectivity (TR$\Delta$R) and transient ellipticity (TRKE) measurements are performed in which the probe laser wavelength is tuned across the energy range of $X_A$ with the pump energy set at 1958 meV. Figure \ref{fig:4}b and c are false-color representations of TR$\Delta$R obtained from domain I and II of Sample\,1 (cf. Fig \ref{fig:1}), respectively, as a function of pump-probe delay. The energy-dependent temporal evolution of reflectivity in individual domains clearly corresponds to their respective steady-state response (superimposed white lines).  For both the domains, TR$\Delta$R traces at the higher resonance (marked by blue arrows) evolve similarly viz. an ultrafast decay component followed by relatively slower dynamics, as depicted by the blue colored traces in Fig. \ref{fig:4}d and e. The response at the lower resonance energy has different characters for different domains. 
For instance, in domain I, we observe a monotonous decrease in the signal from t=0. Remarkably, for domain II, during the initial five ps, the reflectivity of the probe pulse increases with time, followed by a slower decay. The initial increase likely stems from slow state filling at the probe energy during relaxation, possibly suggesting interlayer charge transfer prompted by interlayer hybridization. These observations, therefore, further substantiate the theoretical understanding developed in the context of steady-state optical response. 

Unlike the transient reflection, the ellipticity signal manifests no prominent energy dependence. Instead, it is remarkably sensitive to the ferroelectric stacking configuration and exhibits entirely different traits for the two domains investigated. Figure \ref{fig:4} f and g portrays the TRKE results for domain I and II probed at higher and lower resonance energies (SI Fig. S28 for the complete spectrum). In domain I, the TRKE decay is relatively slow, albeit faster than the decay of the transient reflectivity traces, indicating that it is not predominantly driven by exciton recombination but by a spin dephasing mechanism on a similar timescale as observed in MoS$_2$ monolayers\cite{UrbaszekMariePRL2014, GlazovPRb2014}.   By contrast, in domain II, the excitons lose their spin polarization almost instantly. The microscopic mechanism for this striking dependence of exciton lifetimes and spin-valley dynamics on ferroelectric order remains an open question. However, recent calculations\cite{jafari2023robust,frank2024emergence} show that in parallel-stacked TMDs, dependent on the symmetry of the stacking order, in-plane spin-orbit coupling terms can arise that lead to spin mixing of K states. In this way, specific stacking orders may facilitate spin dephasing, while others protect spin orientation, yielding prolonged spin-valley lifetimes comparable to TMD monolayers. This opens up the use of spin and valley as useful degrees of freedom in 3R-stacked multilayers, which have so far been limited predominantly to monolayer TMDs.\\

\subsection*{Conclusions} 
To conclude, we demonstrate the direct correspondence between ferroelectric order and intralayer excitons in 3R-stacked MoS$_2$ multilayers, allowing us to map ferroelectric domains directly using far-field optical spectroscopy. Ab initio theory provides a comprehensive explanation of the microscopic mechanism. In a field-effect device, we are able to control and map domain wall motion at room temperature and observe clear hysteretic features with a coercive field of roughly 0.03-0.035 V\,nm$^{-1}$. The relatively small switching field of ferroelectricity by domain wall sliding could facilitate efficient electrical control of various optical properties of 2D materials \cite{menghaoWu2023}. Addressing individual domains in time-resolved experiments, we find that depending on ferroelectric order, 3R-stacked MoS$_2$ multilayers can show spin-valley dynamics on timescales comparable to TMD monolayers or near-instantaneous loss of valley polarization. Our study paves the way for utilizing ferroelectric order in few-layer TMDs embedded within functional van der Waals heterostructures as a local, nonvolatile control lever for excitonic and valleytronic properties. Theses findings introduce a distinctive perspective to the field of quantum optoelectronics with van der Waals materials and will enable novel optoelectronic device architectures that rely on efficient electrical manipulation of excitons or the optical response in general.

\subsection*{Methods}
\textbf{Device fabrication:} 3R-MoS$_2$, obtained from HQ Graphene, were exfoliated onto polydimethylsiloxane (PDMS). MoS$_2$ flakes were selected according to their optical contrast. Chosen TMD flakes were stamped on pre-exfoliated \textit{h}BN flakes. We use a polycarbonate (PC)/PDMS-based hot pickup method to transfer \textit{h}BN flakes from bare Si/SiO$_2$ substrate to encapsulate \textit{h}BN/TMD stack. A few layer thick graphite flakes (procured from NGS Naturgraphit), exfoliated on PDMS, are deterministically stamped to make external contact to the sample.

\textbf{KPFM measurements:}
KPFM measurements were performed using Park System NX20. The electrostatic signal was measured at side-band frequencies using two built-in lock-in amplifiers. We used PointProbe Plus Electrostatic Force Microscopy (PPP-EFM) n-doped tips. For the unencapsulated sample (Sample 1 c.f. Fig. 1) we used non-contact mode of operation. The average height above the surface was controlled via a two-pass measurement. The first pass records the topography, whereas, in the second pass, the tip follows the same scan line with a predefined lift (typically 4-5\,nm) and measures the KPFM signal. In case of Sample 4 c.f. Fig. 3 where we used a $\sim$85 nm \textit{h}BN for top encapsulation, we operated the AFM in tapping mode and the KPFM measurements were done in a single pass.

\textbf{Reflectance  measurements and Spatial filtering:}
For reflectance measurements, the samples were illuminated with a quartz tungsten halogen lamp. The collimated beam was focused on the sample using an 80X objective (NA=0.50, f=200). The reflected light from the sample was collected using the same objective and spectrally resolved using a spectrometer and a charge-coupled detector. In the detection path, before the spectrograph, we introduce a home-built \textit{Spatial-Filtering} module to enhance the spatial resolution. The input side of the spatial filter consists of an aspheric lens mounted on a Z-translator. It focuses the reflected beam onto a pinhole. The pinhole was carefully aligned to the beam path with the help of an XY translator. A fraction of the focused light passes through the pinhole aperture. Another aspheric lens was used to collect the beam from the pinhole and collimate along the detection path.

\textbf{Time resolved reflectivity and Kerr ellipticity:}
In the pump-probe setup, two separately tunable pulsed lasers (Toptica: femtoFiberPro) were used. Each system emitted with a pulse repetition rate of 80 MHz, a spectral width of 6 meV, a pulse duration of about $\sim$200 fs. Probe and pump pulses were electronically time-synchronized with aa auto- and cross-correlation width of $\sim$300 and 700\,fs, respectively. The cross-correlation time of $\sim$700\,fs marks temporal resolution of our measurement. The pump beam was circularly polarized through an achromatic $\lambda/$4 plate. The pump beam was set to a power of 30 $\mu$W and the probe beam was set to 20 $\mu$W. Both the beams were focused with an 80X (NA=0.50, f=200) objective onto the sample. Therefore, the pump and probe fluence roughly correspond to a maximum of 1.7 and 1.1 KW/cm$^2$, respectively. After reflection the pump beam was filtered out by a long-pass filter. The polarization of the probe beam was analyzed for its ellipticity by a combination of a $\lambda$/4 plate, a Wollaston prism, and two photodiodes (ThorLabs PDB210A Si Photodetector). The difference signal of the two diodes was fed into a lock-in amplifier, yielding a TRKE signal. The sum signal of the diodes was fed into a second lock-in amplifier, yielding the TR$\Delta$R signal.

\subsection*{Acknowledgements} 
The authors would like to thank I. Barke for technical assistance. S. D. acknowledges financial support by the Humboldt foundation and a startup funding grant provided by the DFG \emph{via} SPP2244. The authors also acknowledge financial support by the DFG \emph{via} the following grants: SFB1277 (project No. 314695032), SFB1477 (project No. 441234705), SPP2244 projects FA791/8-1 and KO3512/6-1 (project No. 443361515),  INST 264/181-1 FUGG (project No. 441219355) and KO3612/7-1 (project No. 467549803). K. W. and T. T. acknowledge support from the JSPS KAKENHI (grant numbers 21H05233 and 23H02052) and World Premier International Research Center Initiative (WPI), MEXT, Japan.

\subsection*{Author contributions}
S.D. and T.K. conceived the idea for the study. S.D. performed the experiments, prepared the samples together with J.K., analyzed the data together with T.K. and P.F.J.  and wrote the manuscript in close interaction with all other authors. M. K., R. S., and T. K. designed, built, and  tested the setup for the time-resolved Kerr experiments. S.D. upgraded the setup for simultaneous time-resolved reflectivity measurements. K.W. and T.T. provided high-quality \textit{h}BN crystals for sample preparation. P.F.J. and J. F. performed the DFT calculations and analyzed the results together with S.D. and T.K. 

\subsection*{Competing interests}
The authors declare no competing interests.

\newpage

\section*{Supplementary Information}
\section{Device fabrication}
\textit{h}BN flakes of various thicknesses were mechanically exfoliated onto a Si/SiO$_2$ substrate which were annealed on a hot-plate at 360\textcelsius to clean up scoth tape residue. 3R-MoS$_2$, obtained from HQ Graphene, were exfoliated onto polydimethylsiloxane (PDMS). MoS$_2$ flakes were selected according to their optical contrast. Chosen TMD flakes were stamped on the \textit{h}BN surface directly from PDMS at ambient conditions. We use a polycarbonate (PC)/PDMS-based hot pickup method \cite{zomer2014fast} to transfer \textit{h}BN flakes from bare Si/SiO$_2$ substrate to encapsulate \textit{h}BN/TMD stack.

A few layer thick graphite flakes (procured from NGS Naturgraphit), exfoliated on PDMS, are deterministically stamped to make external contact to the sample. Subsequently we wire bond the graphite flake using two component conductive silver epoxy. Keithley 2401 source meter were employed to apply external gate voltage.

\section{KPFM measurements}
KPFM measurements were performed using Park System NX20 AFM either in tapping or in non-contact scanning mode. The electrostatic signal was measured at side-band frequencies using two built-in lock-in amplifiers. We used PointProbe Plus Electrostatic Force Microscopy (PPP-EFM) n-doped tips with a conductive coating  acquired from Nano and More GMBH. The mechanical resonance frequency of the tips was $\sim$75\,kHz, and the force constant was 3\,N/m. The cantilever was excited with an AC voltage to perform KPFM measurements, with an amplitude of 1.5-2.5\,V and a frequency of 2-4\,kHz. In the closed-loop measurements, the DC voltage was controlled by a bias servo to obtain the surface potential.

For unencapsulated sample (Sample 1 c.f. Fig. 1) we used non-contact mode of operation. The average height above the surface was controlled via a two-pass measurement. The first pass records the topography, whereas, in the second pass, the tip follows the same scan line with a predefined lift (typically 4-5\,nm) and measures the KPFM signal. In case of Sample 4 c.f. Fig. 3 where we used a $\sim$85 nm \textit{h}BN for top encapsulation, we operated the AFM in tapping mode and the KPFM measurements were done in a single pass. Images were acquired using the Park SmartScan software, and the data was analyzed with the Gwyddion software.

\section{Reflectance  measurements and spatial filtering}

All optical measurements were performed in flow-type cryostats with an accessible temperature range of 4K to room temperature. The samples were kept under vacuum for all the optical measurements. All the experiments were performed in back-scattering geometry.

For white light reflectance measurements, the samples were illuminated with a quartz tungsten halogen lamp. The collimated beam was focused on the sample using an 80X objective. The reflected light from the sample was collected using the same objective and spectrally resolved using a spectrometer and a charge-coupled detector. To obtain the hyperspectral map of the samples, the cryostat with the samples inside, was moved with respect to the fixed optical spot using a computer-controlled XY stage. The data was analyzed with LabView and Origin software.

In the detection path, before the spectrograph, we introduce a home-built \textit{Spatial-Filtering} module to enhance the spatial resolution. The input side of the spatial filter consists of an aspheric lens mounted on a Z-translator. It focuses the reflected beam onto a pinhole. The pinhole was carefully aligned to the beam path with the help of an XY translator. A fraction of the focused light passes through the pinhole aperture. Another aspheric lens was used to collect the beam from the pinhole and collimate along the detection path.

\section{Time resolved reflectivity and Kerr ellipticity}
In the pump-probe setup, two separately tunable pulsed lasers (Toptica: femtoFiberPro) were used, one for the pump and the other for the probe laser beam. Each system emitted with a pulse repetition rate of 80 MHz, a spectral width of 6 meV, a pulse duration of about $\sim$200 fs. Probe and pump pulses were electronically time-synchronized and amplitude-modulated with different chopping frequencies, adding up to a sum modulation frequency of $\sim$950 Hz. A mechanical delay line changed the path length of both beams and, thus, the time offset of both pulses. Through an achromatic $\lambda/$4 plate, the pump beam was circularly polarized to either left- or right-handed helicity. After reflection the pump beam was filtered out by a long-pass filter while the polarization and signal amplitude of the probe beam were measured. The polarization of the probe beam was analyzed for its ellipticity by a combination of a $\lambda$/4 plate, a Wollaston prism, and two photodiodes (ThorLabs PDB210A Si Photodetector). The difference signal of the two diodes was fed into a lock-in amplifier, yielding a TRKE signal. The sum signal of the diodes was obtained by using an external adder and fed into a second lock-in amplifier, yielding the TR$\Delta$R signal. Both lock-in amplifiers were fed the same reference frequency given by the modulation sum frequency. With this approach, it was possible to measure the TRKE and TR$\Delta$R at the same time, cutting the needed measurement time in half, compared to, as commonly performed, subsequent measurements. The TRKE setup is described in detail elsewhere\cite{Kempf2021, kempf2023rapid}.

\section{Density functional theory}

To investigate the electronic properties and optical selection rules of the five-layer (5L) MoS$_2$ with 3R stacking, we performed density functional theory (DFT) calculations using the all-electron full-potential linearized augmented plane-wave (LAPW) method implemented in the Wien2k code\cite{wien2k}. We consider the Perdew-Burke-Ernzerhof\cite{Perdew1996PRL} exchange-correlation functional with van der Waals interactions included via the D3 correction\cite{Grimme2010JCP}. The wave function expansion into atomic spheres uses orbital quantum numbers up to 10 and the plane-wave cut-off multiplied with the smallest atomic radii is set to 8. Spin–orbit coupling was included fully relativistically for core electrons, while valence electrons were treated within a second-variational procedure\cite{Singh2006} with the scalar-relativistic wave functions calculated in an energy window up to 2 Ry. Self-consistency was achieved using a two-dimensional Monkhorst–Pack k-grid with 15$\times$15$\times$1 points and a convergence criteria of $10^{-6} \; e$ for the charge and $10^{-6} \; \textrm{Ry}$ for the energy. We used the structural parameters taken from Ref.~\citeonline{Kim2021PRB}, i. e., the in-plane lattice parameter is 3.191 $\textrm{\AA}$, the thickness of a single MoS$_2$ layer is 3.116 $\textrm{\AA}$, the interlayer distance between adjacent MoS$_2$ layers is 3.094 $\textrm{\AA}$, and a vacuum region of 20 $\textrm{\AA}$ was used to avoid interaction between the heterostructures' replicas. We explored all the possible combinations of 3R stackings in a 5L system leading to a total of 10 different structures, shown in Fig.~\ref{fig:5L_stackings} (see also Fig.~\ref{fig:S11}). The total energy of these different systems are nearly the same, as shown in Fig.~\ref{fig:total_energy}, supporting the appearance of different domains in the studied samples. The calculated band structures with spin-orbit coupling along the $\Gamma-K-M$ line are shown in Fig.~\ref{fig:bsGKM} with the color-code representing the spin expectation value in the out-of-plane direction. A zoom of the conduction and valence bands in the vicinity of the K point is shown in Fig.~\ref{fig:bszoomK}, revealing that the energy bands that generate A and B excitons show strong spin-valley locking. The wave functions presented in Fig.~2 of the main text were evaluated without spin-orbit coupling. Turning to the selection rules at the K point, we present in Fig.~\ref{fig:pcv_label} the absolute value of the dipole matrix elements ($|p_{cv}| = \frac{\hbar}{m_0} |\left\langle c, K \left| \vec{p} \cdot \hat{\alpha} \right| v, K \right\rangle|$ calculated within the LAPW basis set\cite{Draxl2006CPC}, with $\hat{\alpha}=s^+, s^-, z$ encoding the light polarization) as a function of the band labels. In Fig.~\ref{fig:pcv2_label} we display the oscillator strength, i. e., $|p_{cv}|^2$, that enters the absorption spectra. In Figs.~\ref{fig:pcv_energy} and \ref{fig:pcv2_energy} we present $|p_{cv}|$ and $|p_{cv}|^2$ as a function of the transition energy only for $s^+$ polarization (the dominant contribution due to the stron spin-valley locking). Our results reveal that variation of the transition energies is on the order of 10--20~meV for the different stackings.

\section{Exciton binding energies}

To evaluate the effect of the asymmetric dielectric surroundings of the 5L MoS$_2$ systems, we calculated the binding energies for the intralayer A excitons located at each layer. We employed the effective Bethe-Salpeter equation\cite{Rohlfing1998PRL, Rohlfing2000PRB} formalism, assuming non-interacting parabolic bands for electrons and holes\cite{Ovesen2019CommPhys, FariaJunior2023TDM}. We consider the average value of effective masses taken from the DFT calculations ($m_v = 0.515$  and $m_c = 0.415$) and the electrostatic potential for each layer is obtained numerically by solving the Poisson equation in $k$-space, assuming the different regions to have dielectric constants of $\varepsilon(\textrm{MoS}_2) = 15.45$ (from Ref.~\citeonline{Laturia2018npj2D}), $\varepsilon(\textrm{hBN}) = 4.5$ (from Ref.~\citeonline{Stier2018PRL}), and $\varepsilon(\textrm{air}) = 1$. We consider each MoS$_2$ layer to have an effective thickness of 6.232~$\textrm{\AA}$, taken as twice the value of the physical thickness of the TMDC\cite{Berkelbach2013PRB}. The calculated dielectric constants experienced by the intralayer excitons at each MoS$_2$ layer are show in Fig.~\ref{fig:eps}. The potential at each layer, $l$, takes the form $V(\vec{k}) \sim \left[ k \epsilon_l(k) \right]^{-1}$, with $l=1,\ldots,5$. The resulting exciton binding energies are given in Table~\ref{tab:Eb}, showing variations of up to 30~meV, a similar energy scale of the optical transitions discussed in Figs.~\ref{fig:pcv_energy} and \ref{fig:pcv2_energy}. For the structures with mixed layer contributions (MX-XM-MX-XM, XM-MX-XM-MX, and XM-XM-MX-MX) we average the energy transitions and leave the intralayer exciton potentials intact. Since these quantities (transition energies and exciton binding energies) operate on the similar energy scales, our conclusions remain valid. Thus, combining the DFT transition energies and dielectric effects within the exciton picture, we present the resulting absorption exciton spectra in Fig.~\ref{fig:abs}. We used the same value of the dipole matrix element for all transitions. DFT underestimates the band gap and therefore our calculations are red-shifted by $\sim$480~meV in comparison with the experimental values. Nonetheless, the exciton absorption peaks are 10--30 meV apart depending on the particular stacking (ferroelectric domains), in excellent agreement with the experimental findings shown in Fig.~1d of the main text.

\begin{table}[H]
	\caption{Calculated intralayer exciton binding energies in meV.}
	\begin{center}
		\begin{tabular}{ccccc}
			\hline
			\hline
			L1 & L2 & L3 & L4 & L5\tabularnewline
			\hline
			$-$131.6 & $-$117.5 & $-$116.6 & $-$124.2 & $-$152.2\tabularnewline
			\hline
			\hline
		\end{tabular}
	\end{center}
	\label{tab:Eb}
\end{table}

\section{Extended data}

\renewcommand{\thefigure}{{\small S}\arabic{figure}}
\setcounter{figure}{0}

\begin{figure}[h]
	\centering
	\includegraphics[scale = 0.4]{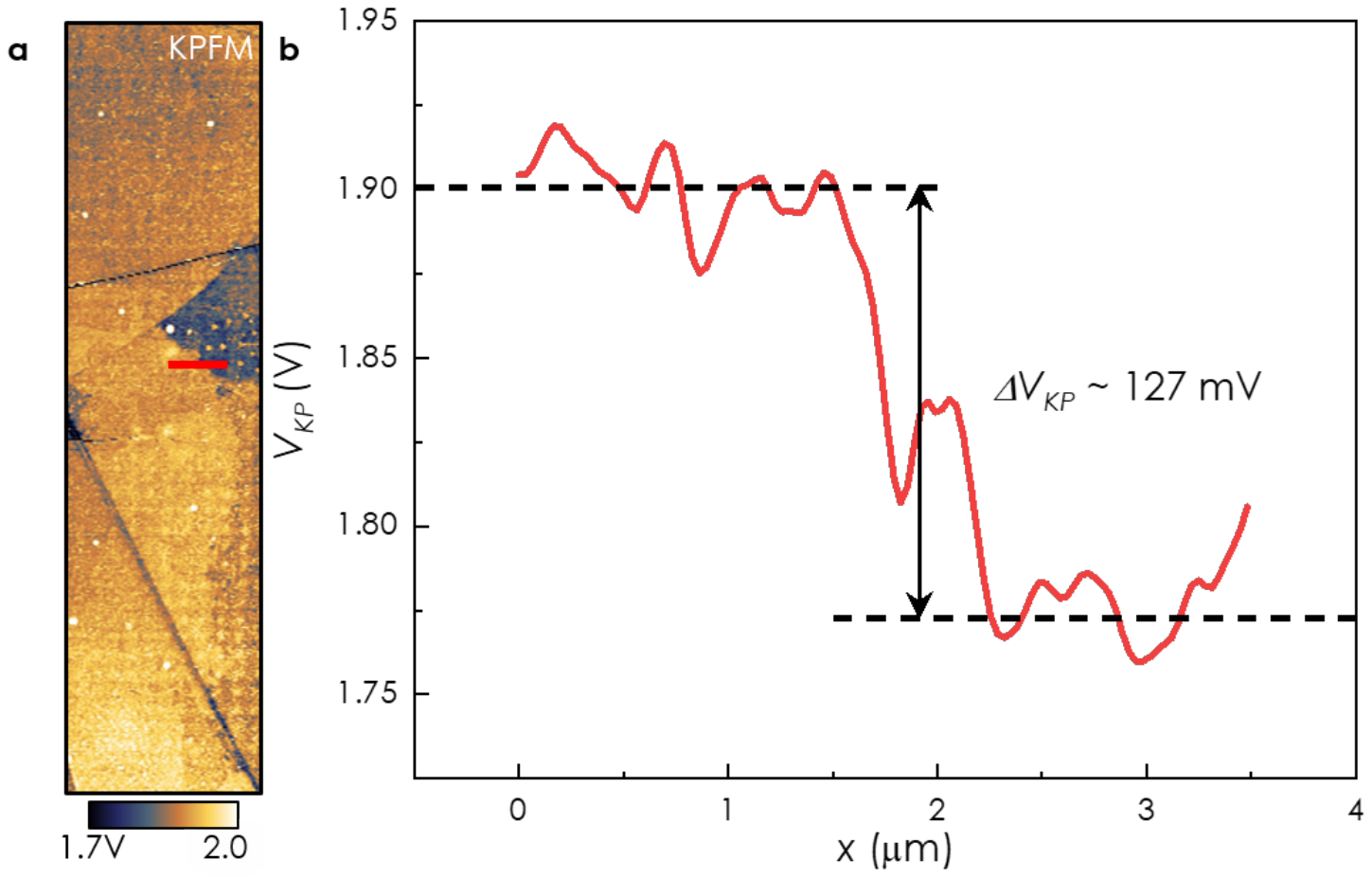}
	\caption{\textbf{Multi-polarization states in naturally grown 3R-MoS$_2$.} \textbf{a.} Surface potential map copied from Fig.1b of the main text for reference. \textbf{b.} Typical line cut of the lateral surface potential profile across a domain wall separating regions I and II, marked by the red line in panel a. The measured potential variation, $\Delta$ $V_{KP}\sim$ 127\,mV agrees well with earlier reported values in this material\cite{deb2022cumulative,cao2024polarization}.}
	\label{fig:S1}
\end{figure}

\begin{figure}[h]
	\centering
	\includegraphics[scale = 0.5]{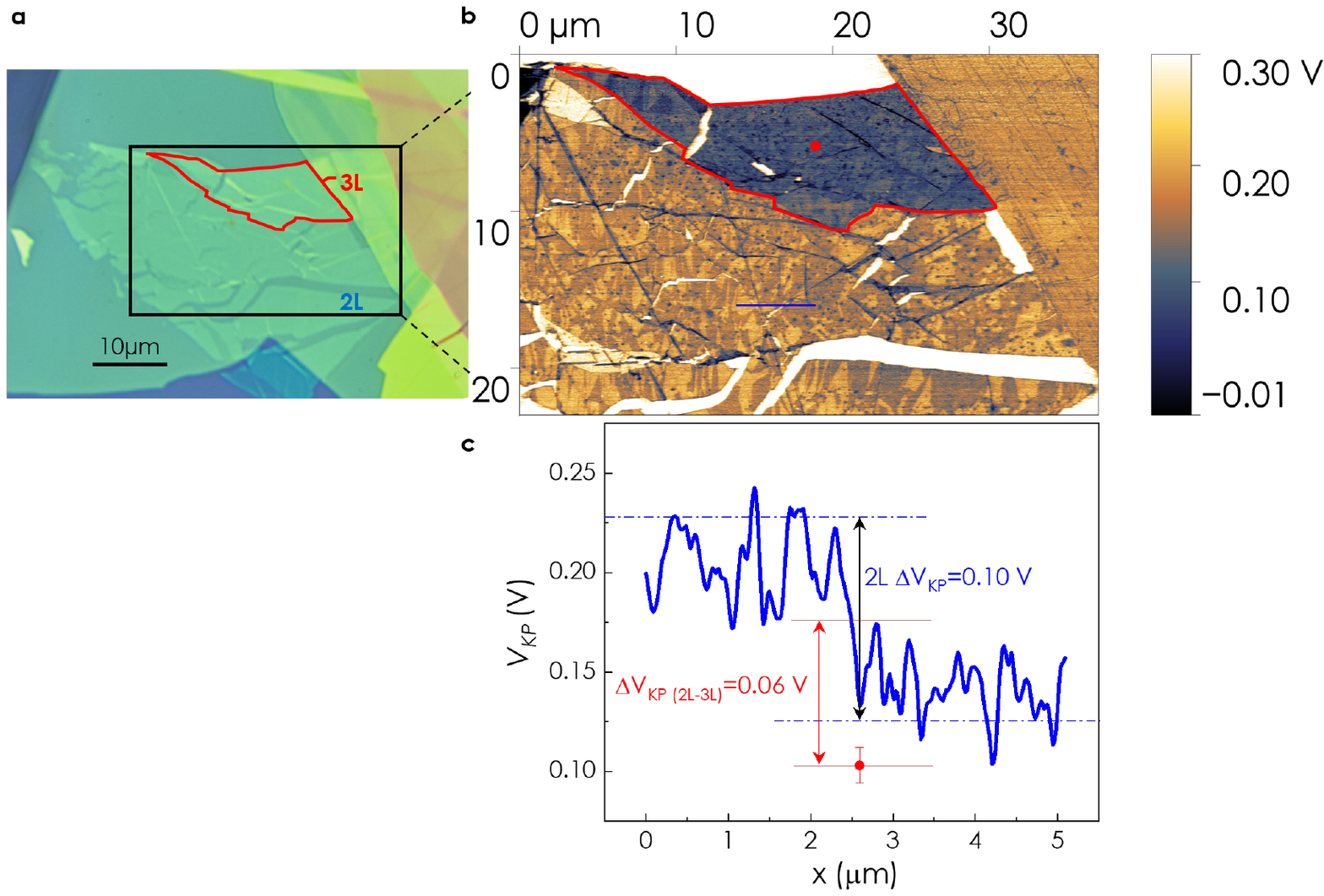}
	\caption{\textbf{Inducing ferroelectric domains in naturally grown 3R-MoS$_2$.} \textbf{a.} Optical micrograph of a bi- and trilayer 3R-MoS$_2$ flake stamped on Si/SiO$_2$/\textit{h}BN with deliberate shear perturbation (horizontal force) during the transfer from PDMS. \textbf{b.} Surface potential map within the area enclosed by the black rectangle in a. Densely packed ferroelectric domains can be identified by the difference in contrast. \textbf{c.} A typical line cut of the lateral surface potential profile across domains in the bilayer region, marked by the blue line in panel b. The red circle is the surface potential from the trilayer region.}
	\label{fig:S2}
\end{figure}

\begin{figure}[h]
	\centering
	\includegraphics[scale = 0.57]{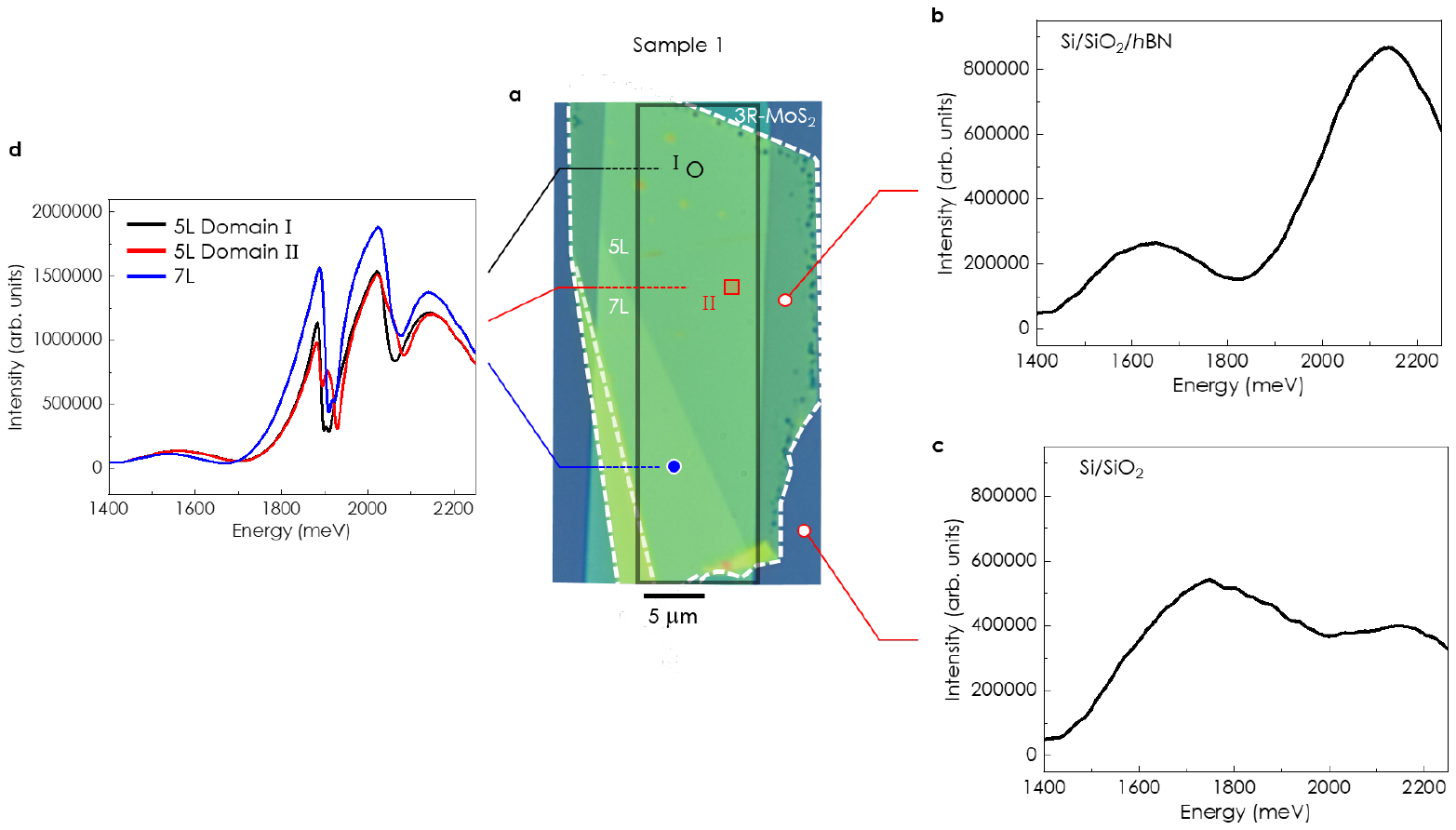}
	\caption{\textbf{Reflectance spectra.} \textbf{a.} Optical micrograph of Sample 1. \textbf{b-d.} As recorded reflectance spectra at 4K from various spatial locations, as indicated by symbols and lines.}
	\label{fig:S3}
\end{figure}

\begin{figure}[h]
	\centering
	\includegraphics[scale = 0.6]{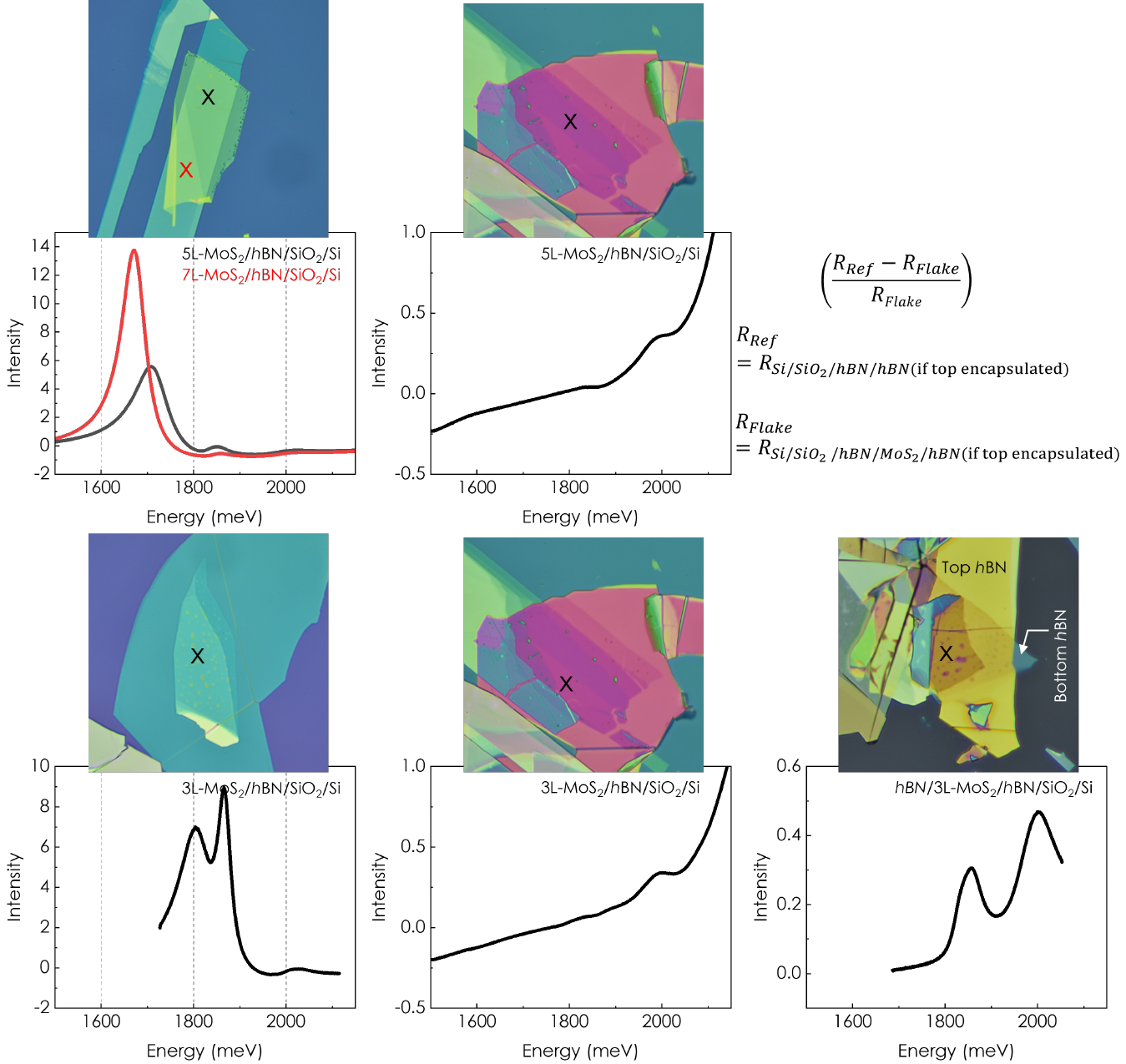}
	\caption{\textbf{Cavity effect on reflection contrast intensity.} Room temperature reflectance contrast spectra of four different samples along with their optical microscopic images. Samples with a turquoise appearance (left panels) shows a pronounced $\Gamma$-Q or K-Q and $X_A$ peaks. Conversely, sample prepared on thicker \textit{h}BN, appearing pink (central panels), exhibit no detectable $\Delta$R/R intensity in the lower energy range, and the $X_A$ transition is only weakly visible. However, the $X_B$ feature is relatively stronger in this sample. In the case of an orange-colored sample (top-encapsulated, right panel), $X_A$ and $X_B$ transitions are prominently present, while the lower energy feature remains invisible. Here we have used RC =  $\Delta$R/R$_{Flake}$.}
	\label{fig:S4}
\end{figure}

\begin{figure}[h]
	\centering
	\includegraphics[scale = 0.6]{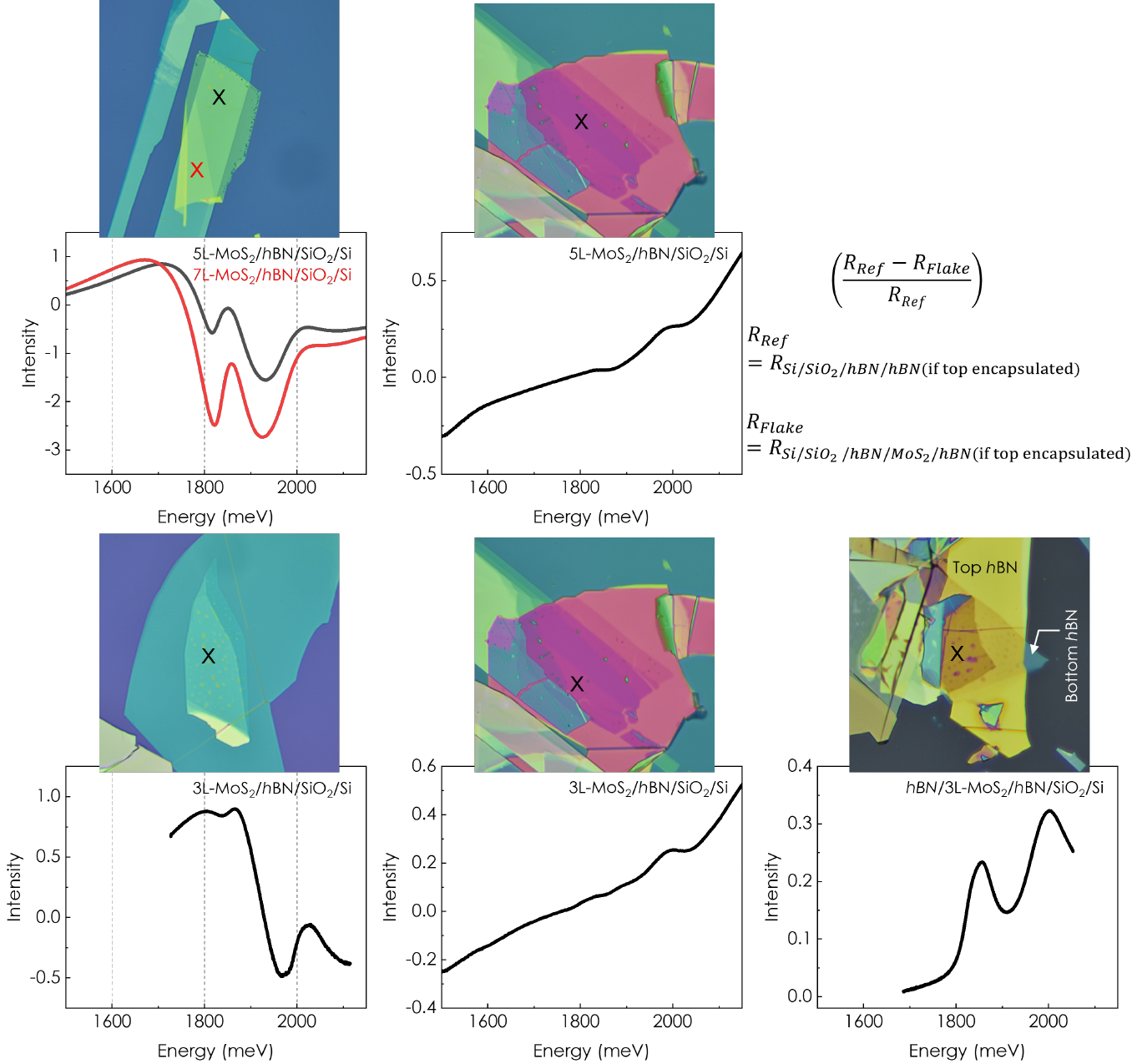}
	\caption{\textbf{Cavity effect on reflection contrast intensity.} The same as in Fig. \ref{fig:S4} with the exception of RC being equal to $\Delta$R/R$_{Ref}$.}
	\label{fig:S5}
\end{figure}

\begin{figure}
	\centering
	\hspace*{-0cm}\includegraphics[scale = 0.7]{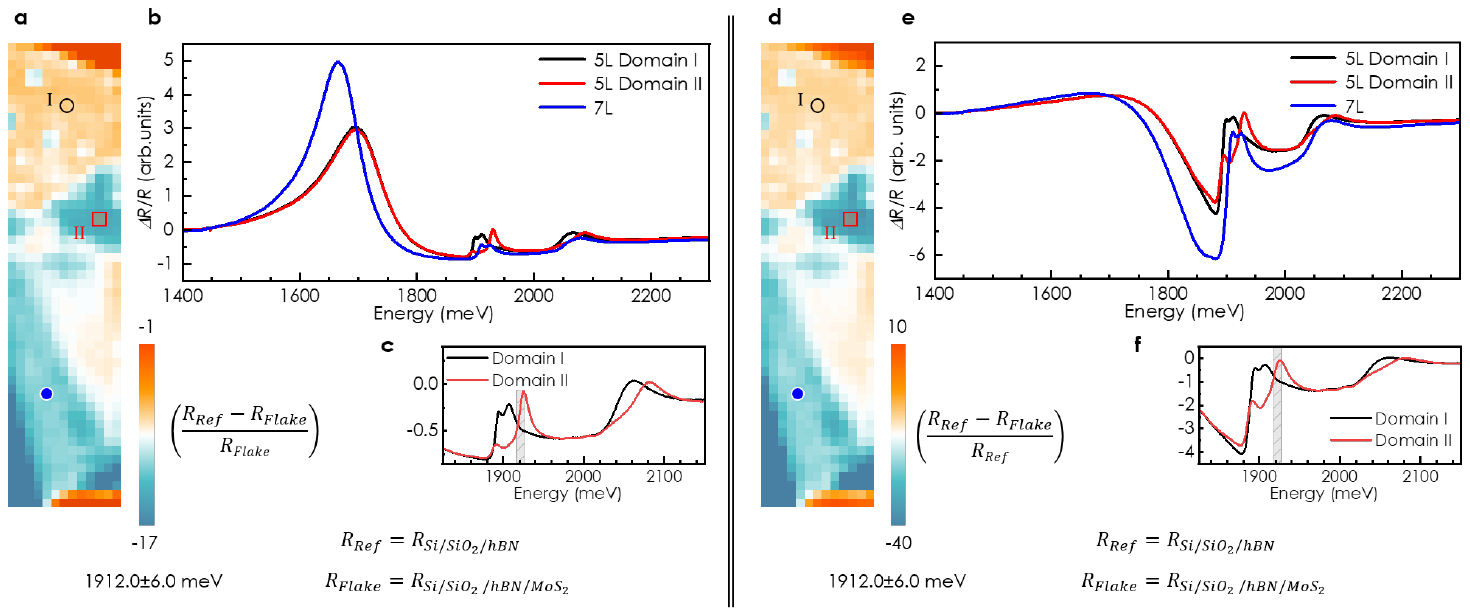}
	\caption{\textbf{Comparison of $\mathbf{(R_{Ref} - R_{Flake})/R_{Flake}}$ with $\mathbf{(R_{Ref} - R_{Flake})/R_{Ref}}$ for Sample 1.} Integrated intensity map of $\Delta R/R$ at the $X_A$ spectral region, i.e., from 1906 to 1918\,meV along with low-temperature reflectance contrast spectra from various spatial locations marked by symbols of corresponding color in the map (black - 5L domain I, red 5L domain II, and blue - 7L). We have used in (\textbf{a-c.}) RC = $(R_{Ref} - R_{Flake})/R_{Flake}$ (as in Fig. 1c and d) and in (\textbf{d-f.}) RC = $(R_{Ref} - R_{Flake})/R_{Ref}$.}
	\label{fig:S10}
\end{figure}

\begin{figure}
	\centering
	\includegraphics[scale = 0.55]{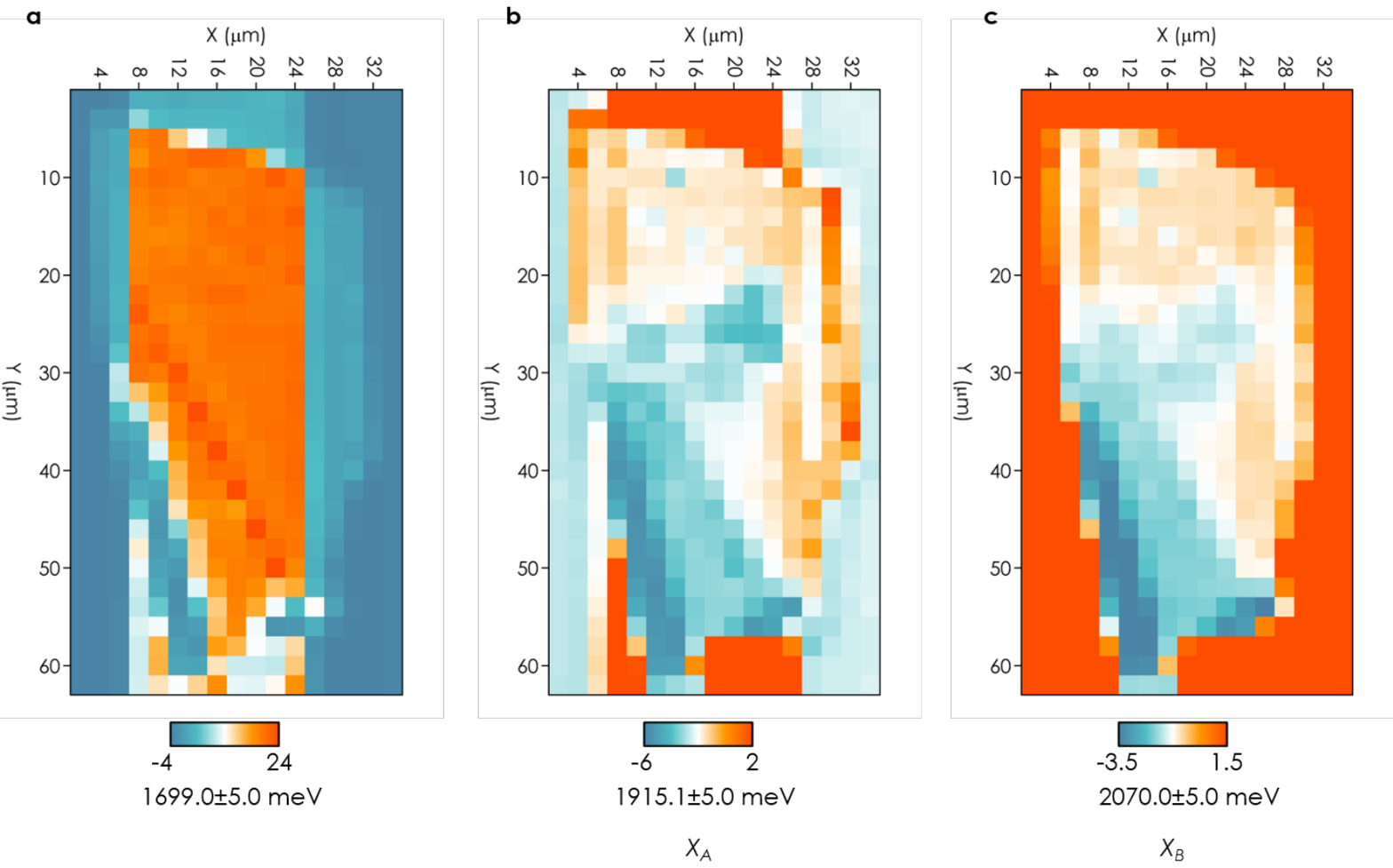}
	\caption{\textbf{Reflection contrast intensity of ferroelectric domains in  few-layer 3R-MoS$_2$.}  \textbf{a-c.} Low-temperature integrated intensity map of $\Delta$R/R of the complete flake at the $\Gamma$-Q or K-Q peak, $X_A$, and $X_B$ spectral region, respectively.% \hl{Experimental conditions: Temperature $\sim$4K, Spectrograph- 300mm, Grating- 150gr./mm. Is it needed?}
	}
	\label{fig:S6}
\end{figure}

\begin{figure}
	\centering
	\includegraphics[scale = 0.5]{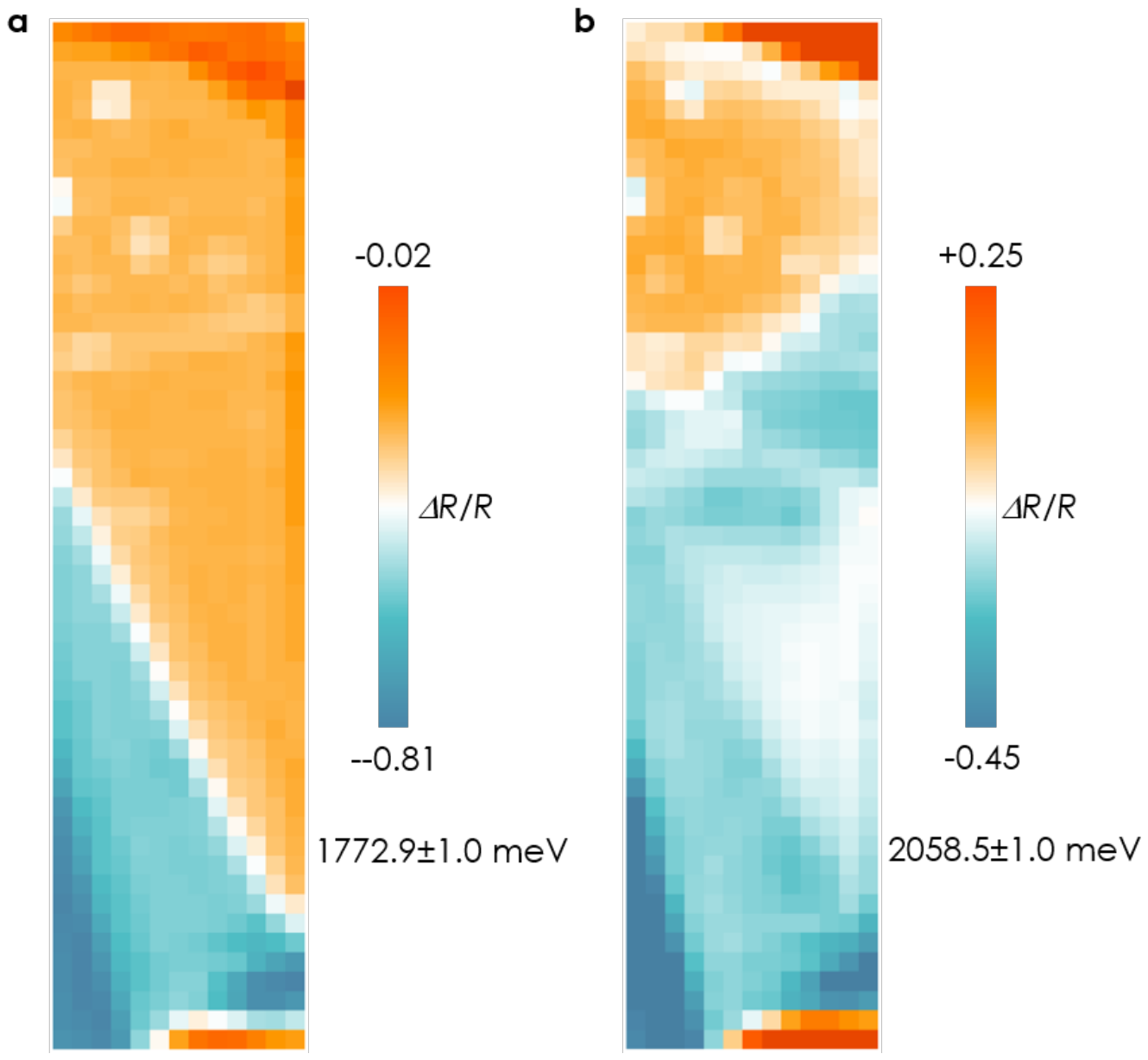}
	\caption{\textbf{High resolution reflection contrast intensity of ferroelectric domains in  few-layer 3R-MoS$_2$.} \textbf{a-b.} Low-temperature integrated intensity map of $\Delta$R/R of the selected region of flake as Fig. 1 in the main text at the low energy tail and B$_{\text{x}}$ spectral region, respectively. This data set differs from Fig. \ref{fig:S6} in terms of its spatial and spectral resolution.% \hl{Experimental conditions: Temperature $\sim$4K, Spectrograph- 300mm, Grating- 600gr./mm. Is it needed?}
	}
	\label{fig:S7}
\end{figure}

\begin{figure}
	\centering
	\includegraphics[scale = 0.6]{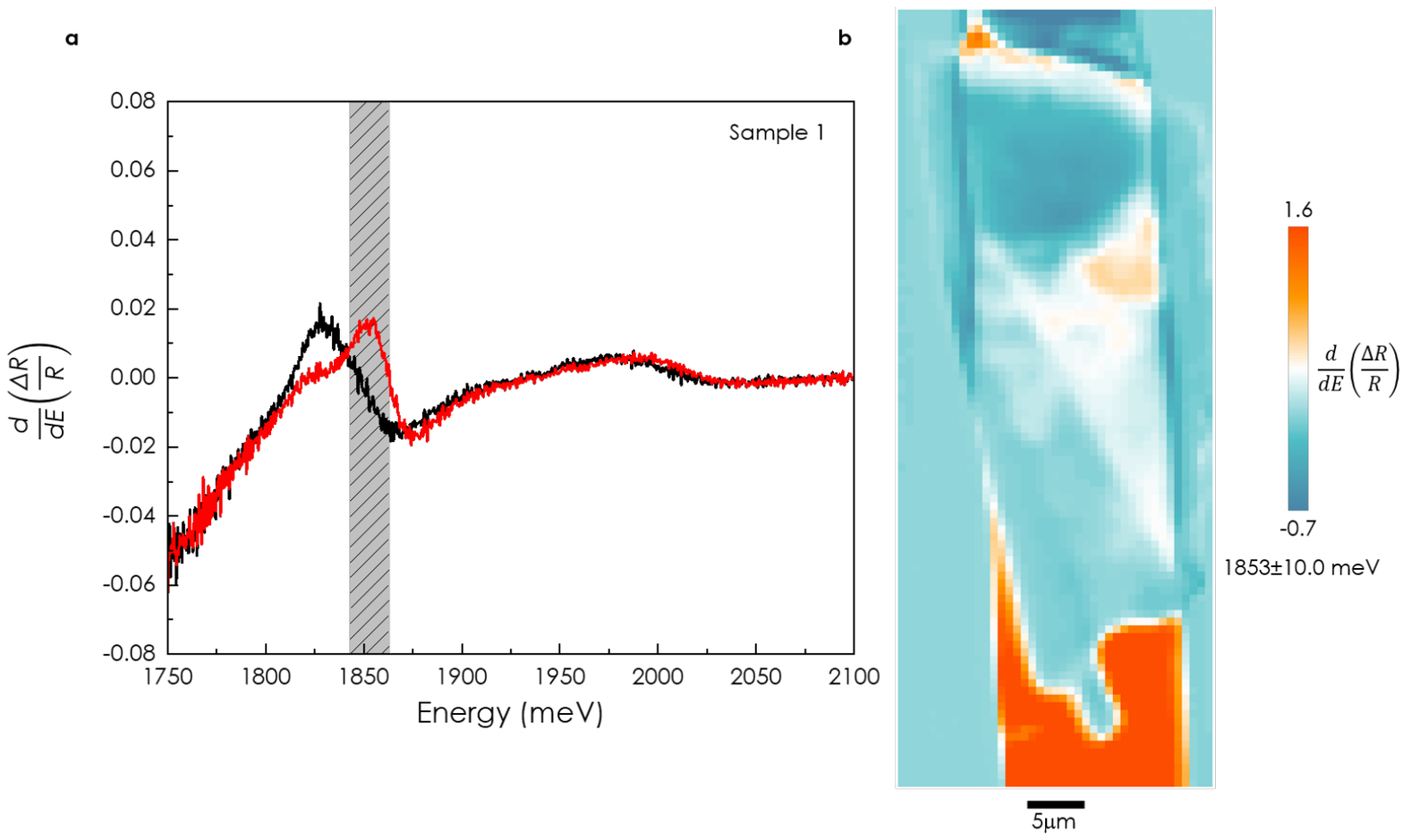}
	\caption{\textbf{First derivative of reflection contrast at room temperature.} \textbf{a.} Numerically calculated first derivative of reflectance contrast spectra from Domain I and II. \textbf{b.} Sum intensity map of $\frac{d}{dE}\left(\frac{\Delta R}{R}\right)$ at the $X_A$ spectral region i.e. from 1843 to 1863\,meV (gray bar in a). % \hl{Experimental conditions: Temperature- Room temp., Spectrograph- 300mm, Grating- 600gr./mm. Is it needed?}
		The analysis highlights that the ferroelectric domains can be probed at room temperature, opening up an unprecedented opportunity for potential applications.}
	\label{fig:S8}
\end{figure}

\begin{figure}
	\centering
	\includegraphics[scale = 0.55]{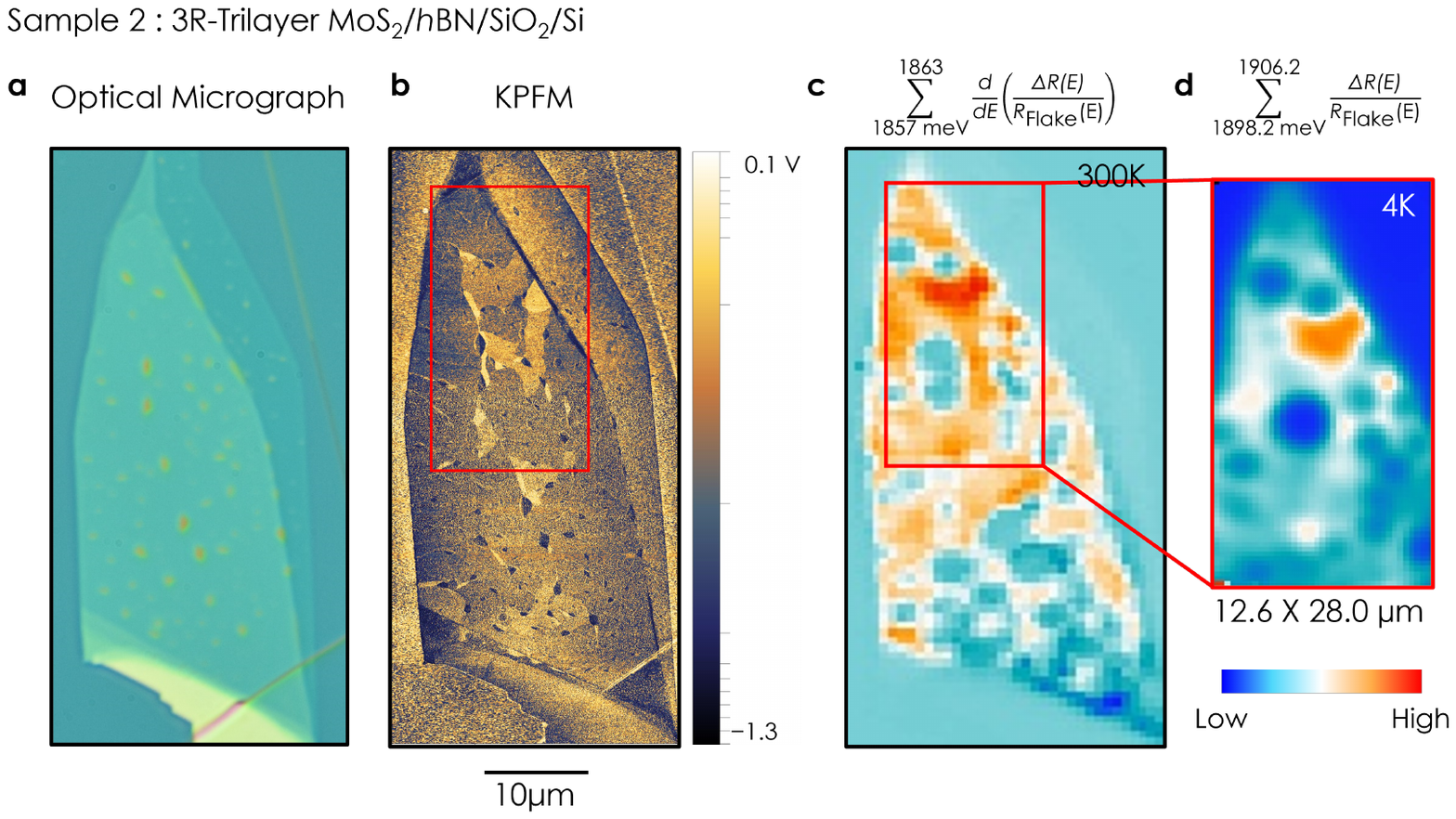}
	\caption{\textbf{Reflection contrast imaging of ferroelectric domains in  trilayer 3R-MoS$_2$.} \textbf{a.} Optical micrograph of a 3R-MoS$_2$ flake on Si/SiO$_2$/\textit{h}BN. \textbf{b.} Surface potential map of the entire flake. \textbf{c.} Room temperature sum intensity map of $\frac{d}{dE}\left(\frac{\Delta R}{R}\right)$ at the $X_A$ spectral region i.e. from 1857 to 1863 meV. \textbf{d.} Low-temperature reflectance contrast map at the $X_A$ spectral region i.e. from 1898.2 to 1906.2 meV.}
	\label{fig:trilayer}
\end{figure}

\begin{figure}
	\centering
	\hspace*{-2cm}\includegraphics[scale = 0.71]{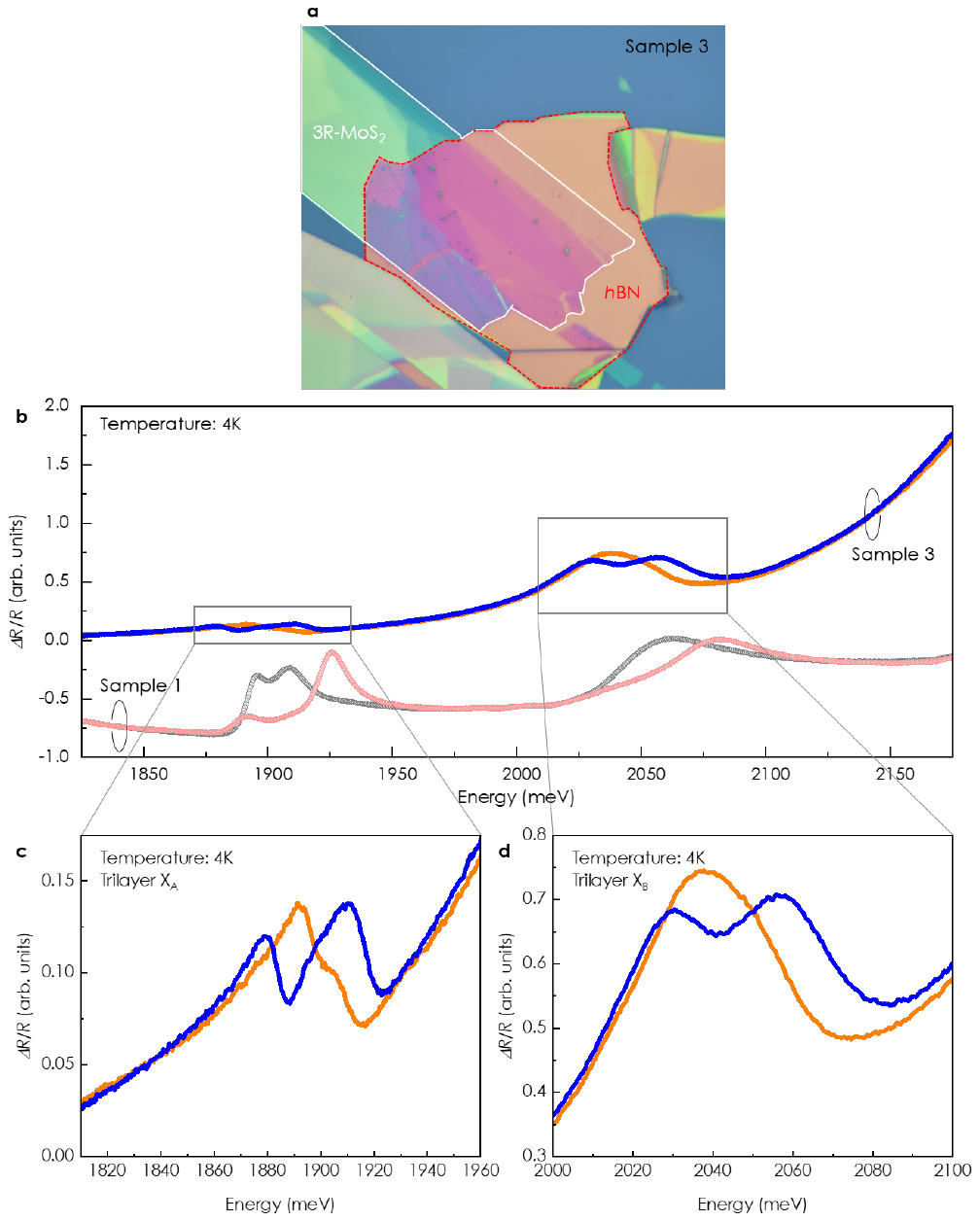}
	\caption{\textbf{Cavity effect on reflection contrast intensity.} \textbf{a.} Another  3R-MoS$_2$ sample on a 95\,nm-\textit{h}BN/285\,nm-SiO$_2$/Si substrate. \textbf{b-d.} Low-temperature reflectance contrast spectra from various spatial locations from the sample shown in panel a, plotted in solid line. Due to the chosen thickness of \textit{h}BN and SiO$_2$ substrate, the reflection contrast at the high energy side increases drastically. The spectra from Fig.1-Sample 1 are included as scatter plots for comparison.% \hl{Data in b in offset corrected. Should we need to mention it?}
	}
	\label{fig:S9}
\end{figure}

\begin{figure}
	\centering
	\hspace{-0.9cm}\includegraphics[scale = 0.6]{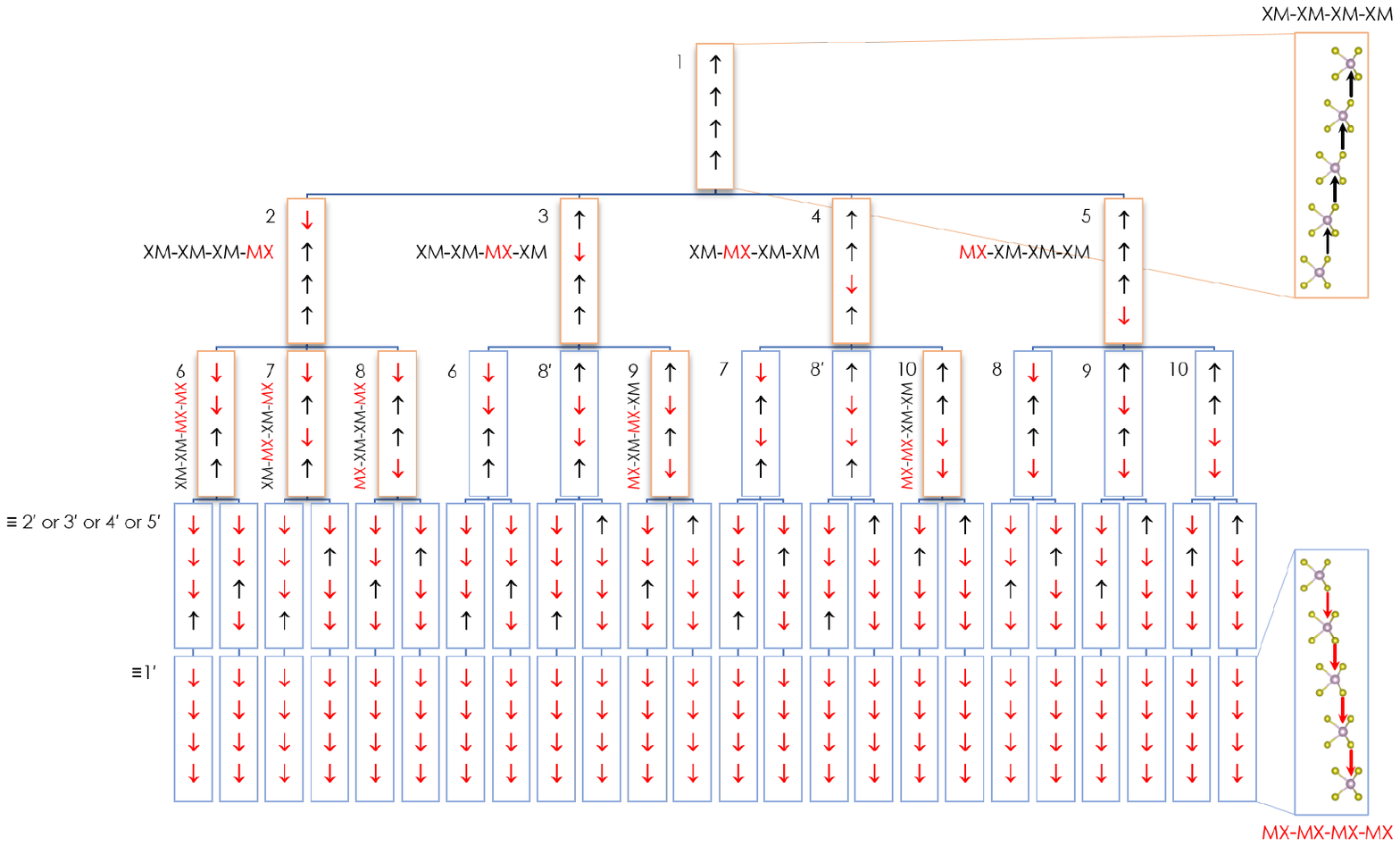}
	\caption{\textbf{Possible permutations of stacking order in a five-layer flake.} We replace the crystallographic depiction with their respective polarization vector for convenience, for instance, an XM stacking configuration with an upwards black colored arrow and MX as a red downwards arrow. Here, we have taken a systematic approach of depicting all the possible combinations by changing the stacking order of one interface at a time, starting from co-polarized XM-XM-XM-XM (1) configuration to entirely flipped MX-MX-MX-MX (1$^{\prime}$) ordering. Orange-colored boxes indicate crystallographically non-equivalent and unique stacking orders. The ``${\prime}$'' symbol has been used to indicate equivalent yet flipped configurations. The scheme shows all sixteen possible stacking combinations, out of which ten are crystallographically unique, and the rest are flipped versions of a few. For example, XM-MX-MX-XM (8$^{\prime}$) is a 180$^o$ flipped version of MX-XM-XM-MX (8). It is interesting to note that KPFM measurement can distinguish among different rows of this schematic representation but is insensitive to the elements within a given row. For example, XM-XM-XM-XM (1) and XM-XM-XM-MX (2) would have different surface potential; however, all the combinations in the second row, from 2 to 5, would remain indistinguishable in a KPFM measurement.}
	\label{fig:S11}
\end{figure}

\begin{figure}
	\centering
	\hspace*{-0.5cm}\includegraphics[scale = 0.54]{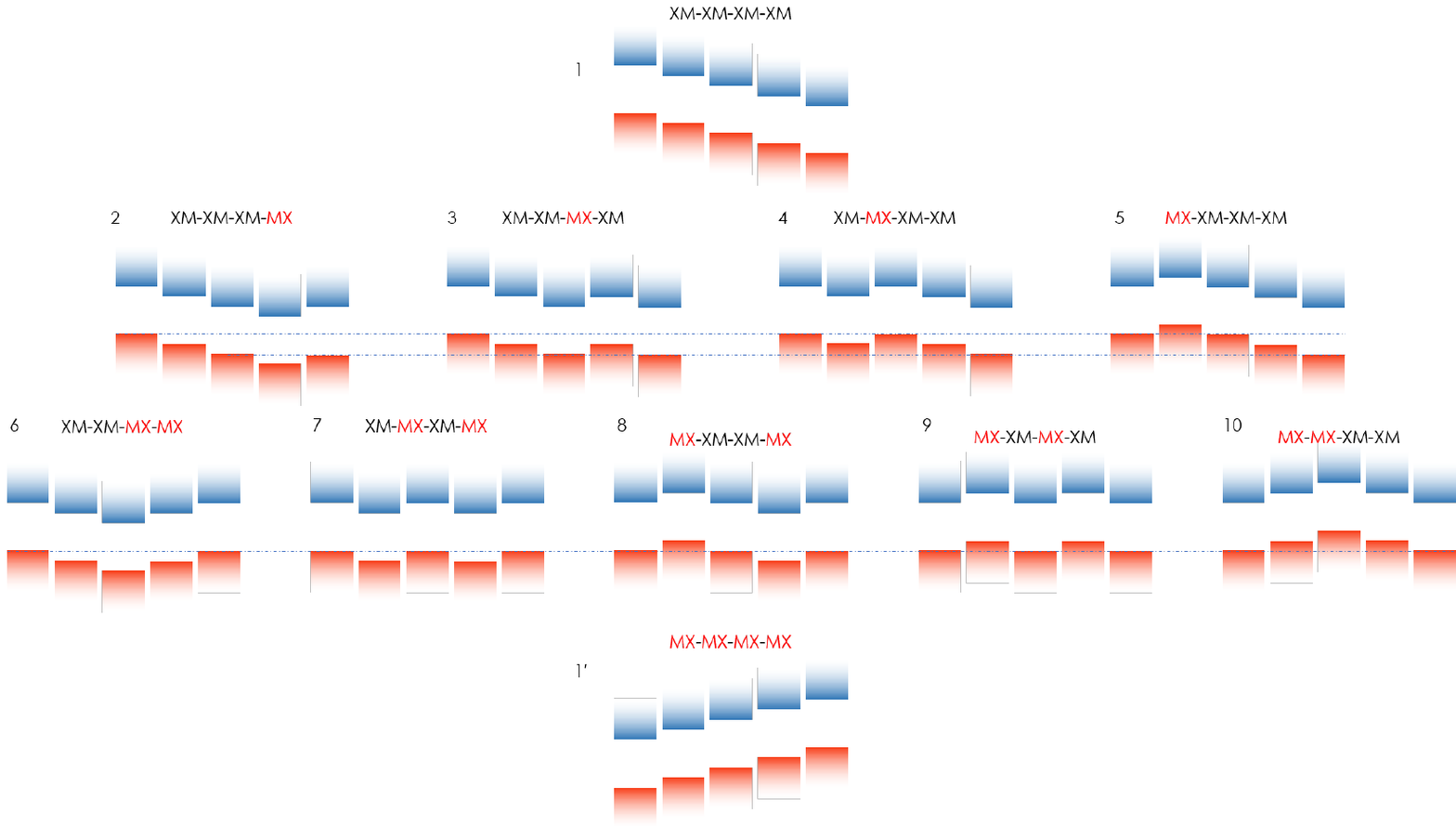}
	\caption{\textbf{Schematic of real-space band alignment for different stacking, a purely electrostatic perspective.} Schematics of layer-projected K$_{\text{VB}}$ and K$_{\text{CB}}$ band-edges for crystallographically unique stacking configurations, discussed in Fig. \ref{fig:S11}. The dashed lines act as a guide to the eye to highlight valence-band degeneracies, which can occur in partially polarized stacking. To first order, the band-edge variation arises due to the rigid shift in layer-projected K valleys introduced by the ferroelectricity-induced electrostatic considerations.}
	\label{fig:S12}
\end{figure}

\begin{figure}
	\centering
	\hspace*{0cm}\includegraphics[scale = 0.6]{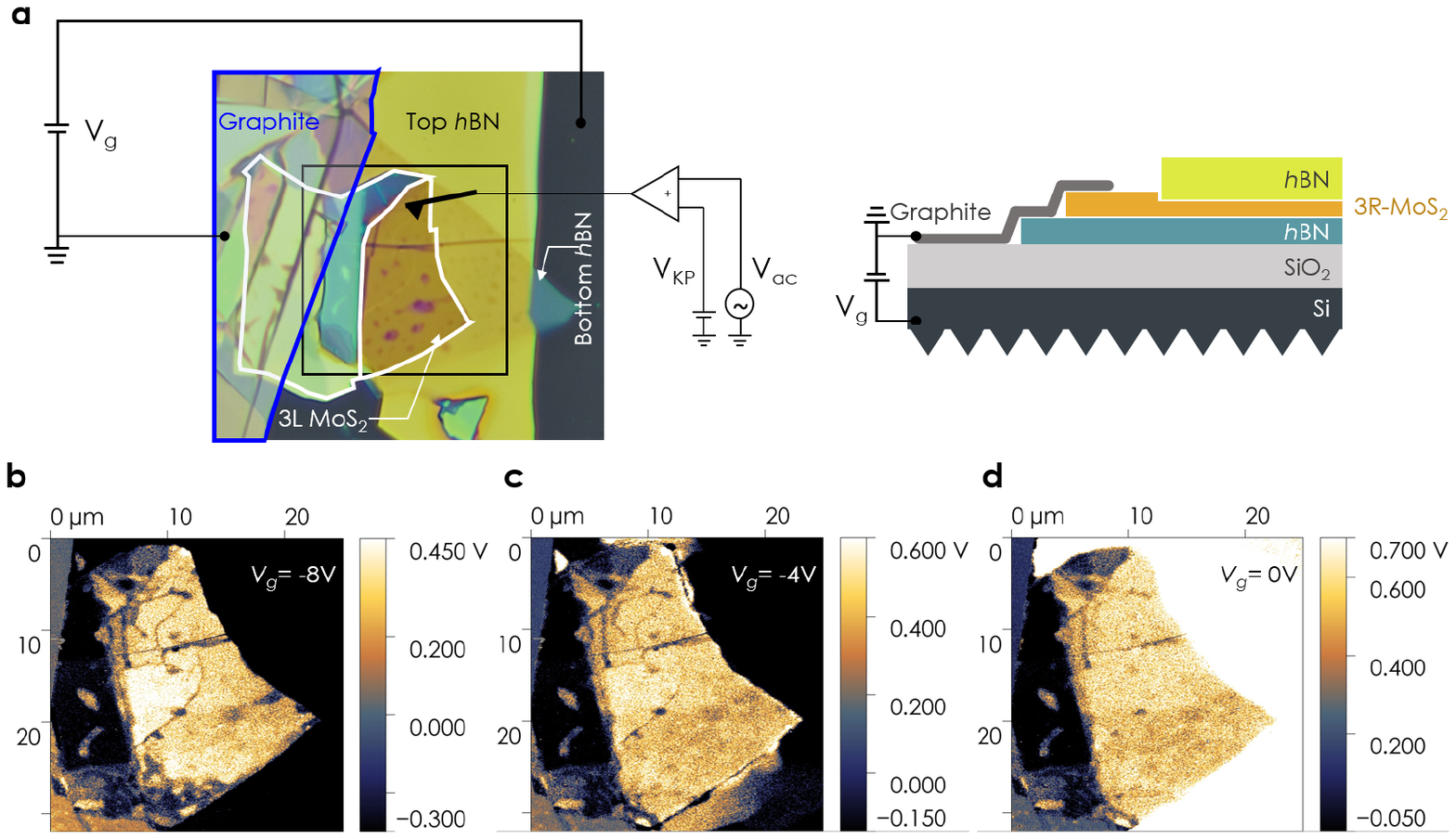}
	\caption{\textbf{Characterization of Sample  4.} \textbf{a.} Optical micrograph of a top-encapsulated 3L-3R-MoS$_2$ flake on Si/SiO$_2$/\textit{h}BN. Topographical steps and edges of the MoS$_2$ flake have been marked by white lines. We have used blue lines to mark the graphite flake used as a source/drain contact. On the right panel we draw the device schematic from a lateral perspective. \textbf{b.} Surface potential map at different gate voltages within the area enclosed by the black rectangle in a. Two domains can be identified by the difference in contrast, which is particularly strong for V$_g$= -8\,V.}
	\label{fig:S13}
\end{figure}

\begin{figure}
	\centering
	\hspace*{0cm}\includegraphics[scale = 0.55]{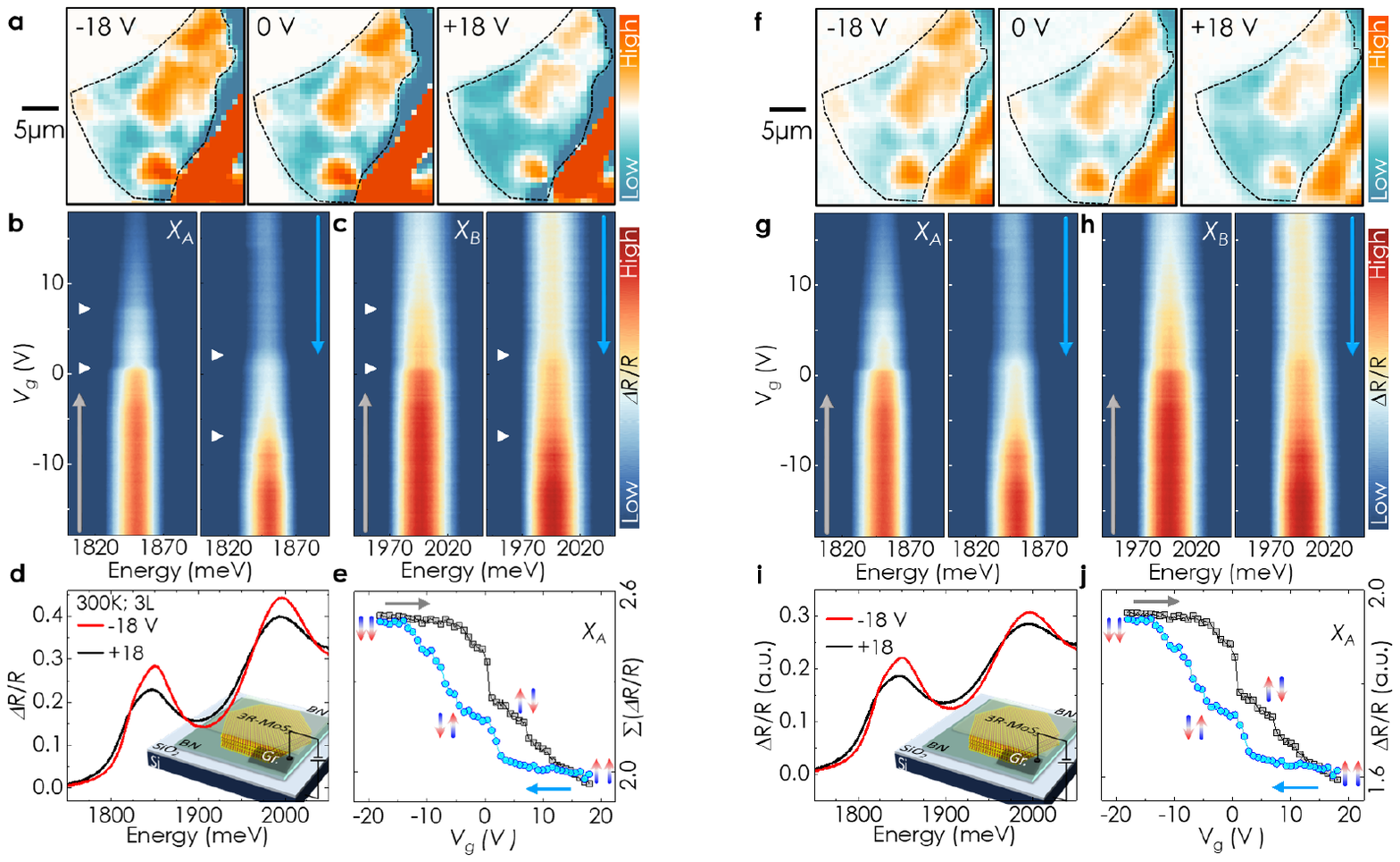}
	\caption{\textbf{Comparison of $\mathbf{(R_{Ref} - R_{Flake})/R_{Flake}}$ with $\mathbf{(R_{Ref} - R_{Flake})/R_{Ref}}$ for Sample 4.} \textbf{a - e} Replica of Fig. 3 i.e. RC maps, RC intensity as a function of gate voltage, RC spectra, and its hysteretic response where we have used RC=$(R_{Ref} - R_{Flake})/R_{Flake}$. \textbf{(f - j)} the same with RC defined as $(R_{Ref} - R_{Flake})/R_{Ref}$,
	}
	\label{fig:S14}
\end{figure}

\begin{figure}
	\centering
	\hspace*{-1cm}\includegraphics[scale = 0.54]{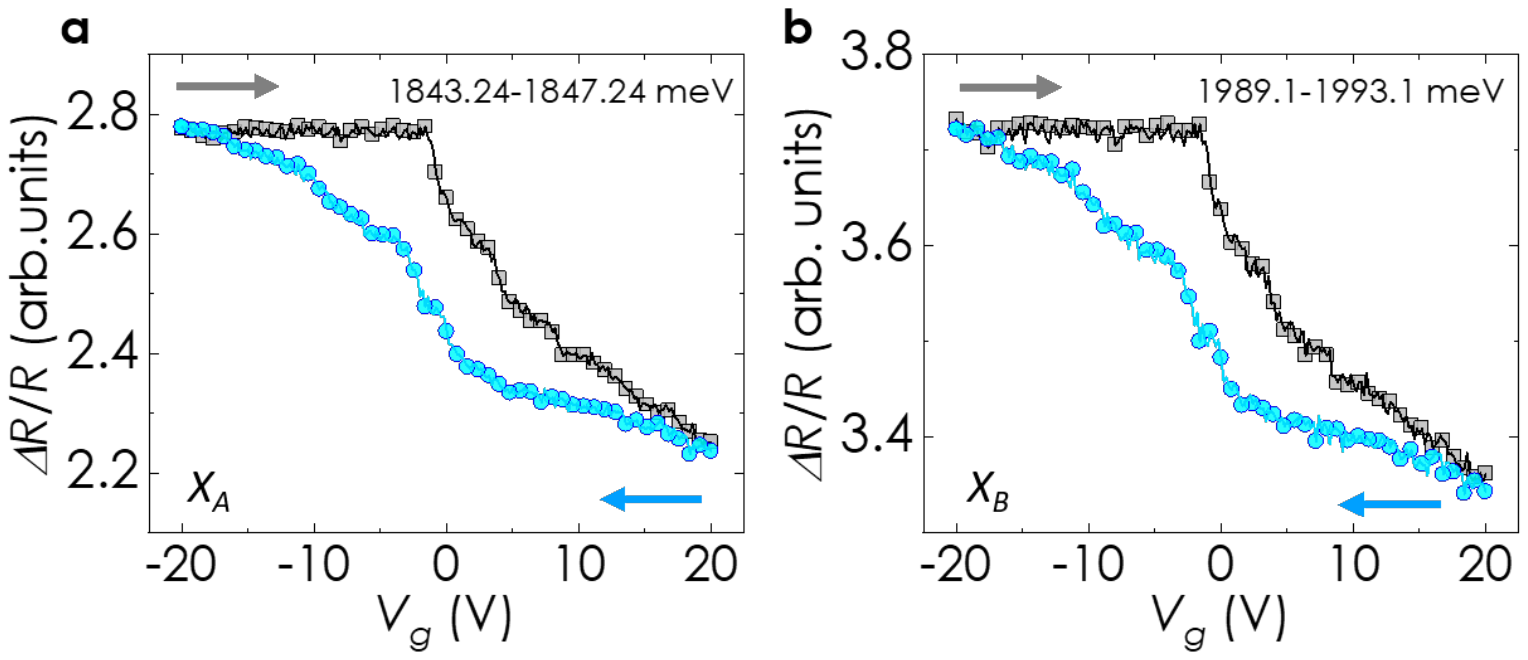}
	\caption{\textbf{ Integrated RC intensity profile as a function of gate voltage at the A and B excitonic region over an extended range.}}
	\label{fig:S16}
\end{figure}

\begin{figure}
	\centering
	\hspace*{-1cm}\includegraphics[scale = 0.54]{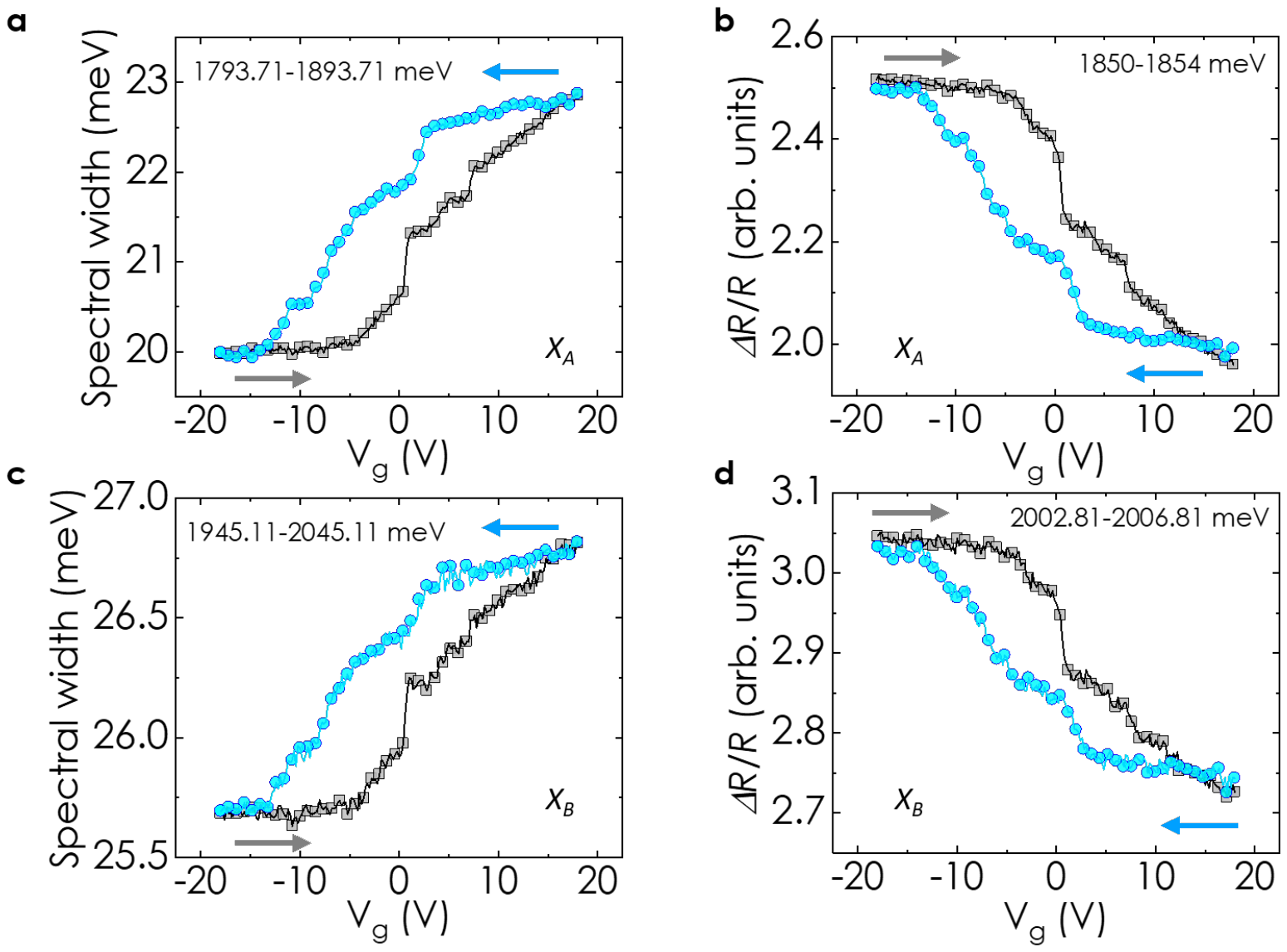}
	\caption{\textbf{Integrated RC intensity ($\equiv$ line cuts from panel b and c of Fig. 3) and standard deviation profile  at the A and B excitonic region.} Integrated intensity =  $\sum_{x=\mathrm{x_{min}}}^{\mathrm{x_{max}}}{\mathrm{\Delta R/R(}x})$; Standard deviation =  $\sqrt{\frac{\sum_{x=\mathrm{x_{min}}}^{\mathrm{x_{max}}}{\mathrm{\Delta R/R(}x)}\left(x-x_0\right)^2}{\sum_{x=\mathrm{x_{min}}}^{\mathrm{x_{max}}}{\mathrm{\Delta R/R(}x})}}-\frac{\sum_{x=\mathrm{x_{min}}}^{\mathrm{x_{max}}}{\mathrm{\Delta R/R(}x)}\left(x-x_0\right)}{\sum_{x=\mathrm{x_{min}}}^{\mathrm{x_{max}}}{\mathrm{\Delta R/R(}x})}$, where $x_0 = (x_{max}+x_{min})/2$, $x$ is the energy of interest. \textbf{a.} Standard deviation with $x_{max} = 1893.71 meV, x_{min} =  1793.71 meV$ as a function of gate voltage. \textbf{b.} Integrated intensity with $x_{max} = 1854 meV, x_{min} =  1850 meV$ as a function of gate voltage (copy of Fig. 3e for completeness). \textbf{c.} Standard deviation with $x_{max} = 2045.11 meV, x_{min} =  1945.11 meV$ as a function of gate voltage. \textbf{d.} Integrated intensity with $x_{max} = 2006.81 meV, x_{min} =  2002.81 meV$ as a function of gate voltage.}
	\label{fig:S15}
\end{figure}

\begin{figure}
	\centering
	\includegraphics{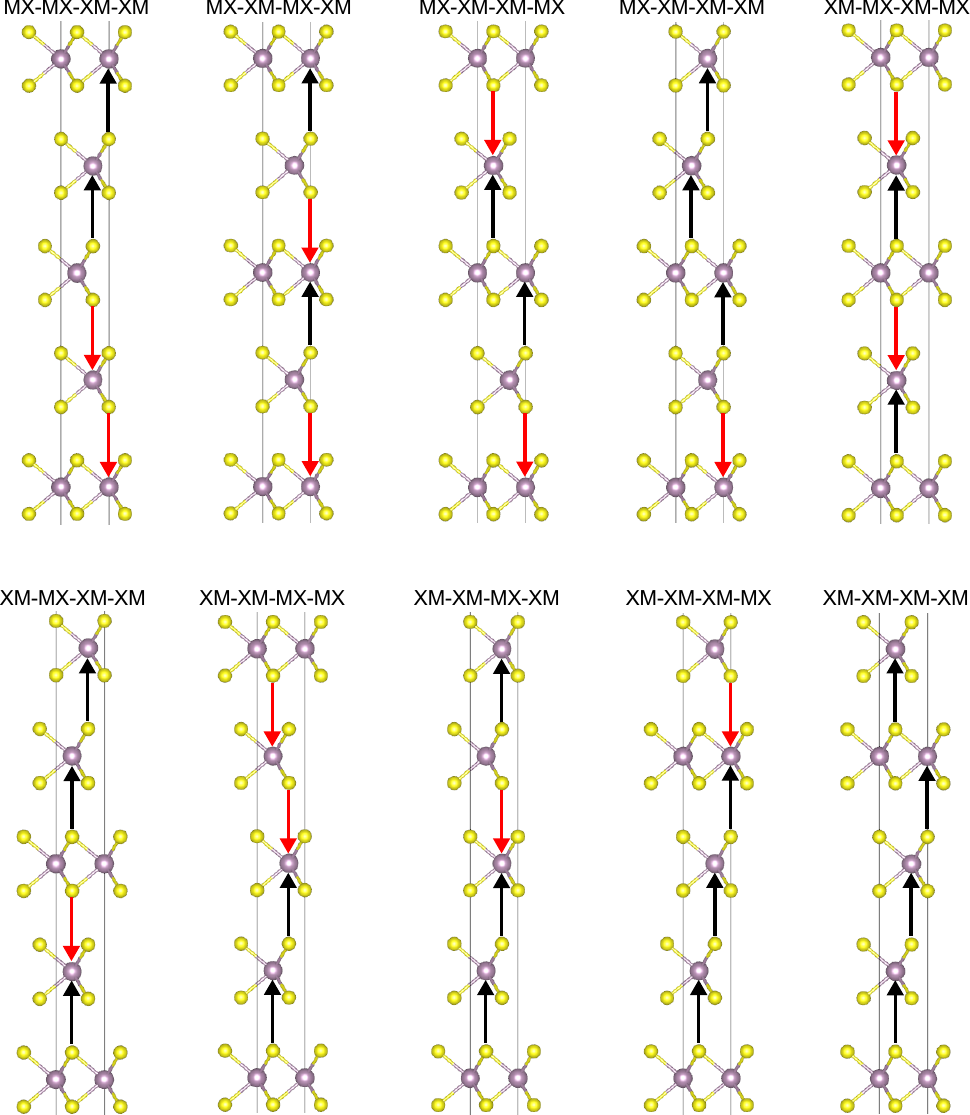}
	\caption{Crystal structures for the different stackings in 3R 5L MoS$_\text{2}$. Red (black) arrows indicate the MX (XM) label.}
	\label{fig:5L_stackings}
\end{figure}

\begin{figure}
	\centering
	\includegraphics{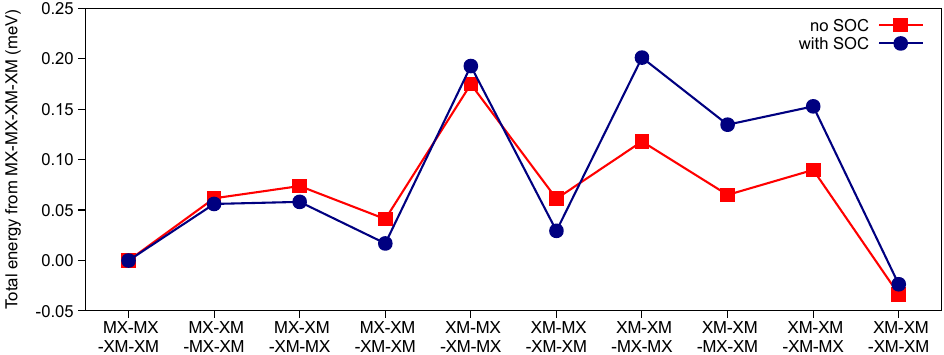}
	\caption{Total energy (with respect to the MX-MX-XM-XM case) for the different stackings.}
	\label{fig:total_energy}
\end{figure}

\begin{figure}
	% \ContinuedFloat
	\begin{subfigure}{\textwidth}
		\centering
		\includegraphics{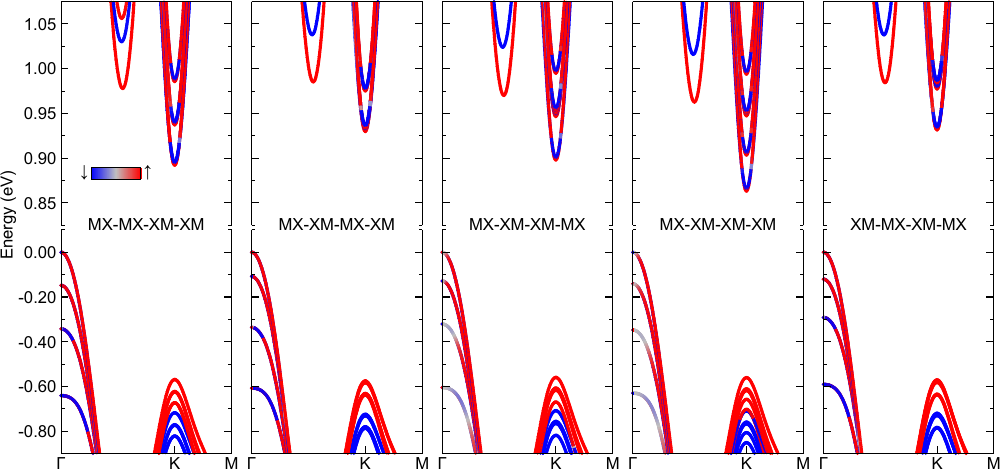}
	\end{subfigure}
	\begin{subfigure}{\textwidth}
		\centering
		\includegraphics{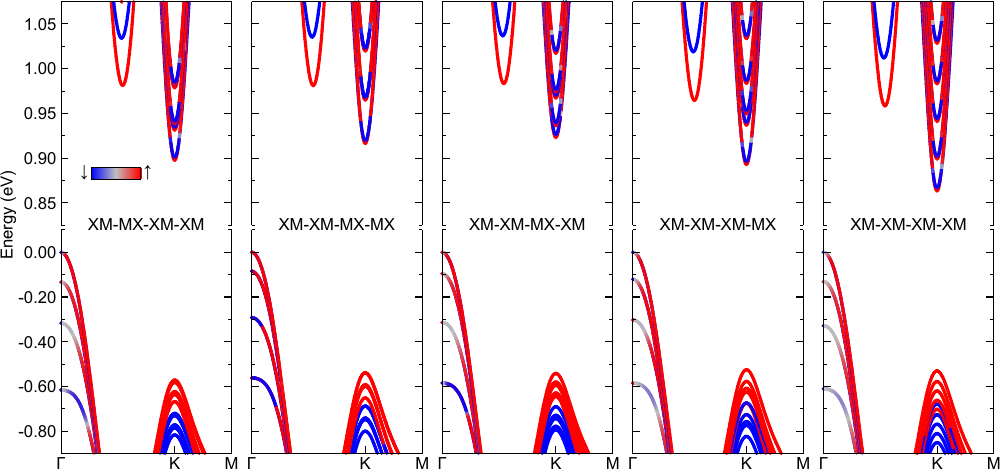}
	\end{subfigure}
	\caption{Band structure along the $\Gamma - K - M$ line, color-coded according to the expectation value of spin along \textit{Z}, $<s_z>$.}
	\label{fig:bsGKM}
\end{figure}

\begin{figure}
	% \ContinuedFloat
	\begin{subfigure}{\textwidth}
		\centering
		\includegraphics{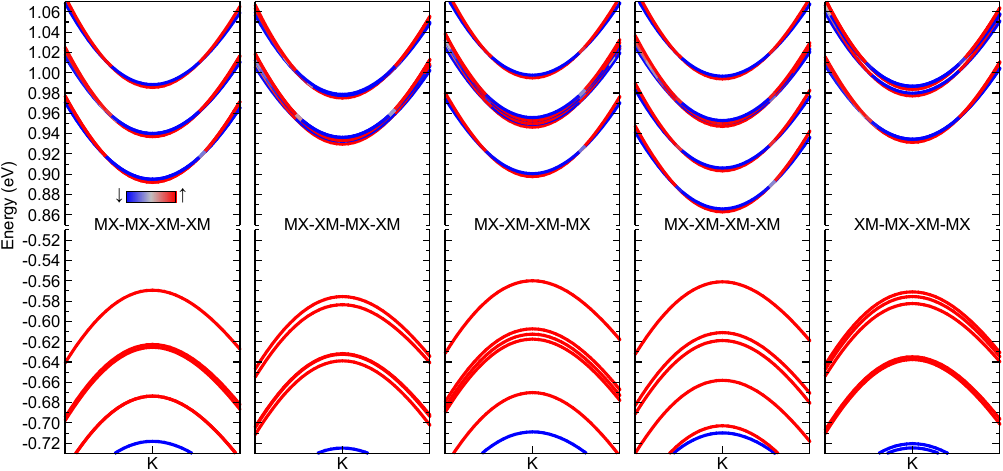}
	\end{subfigure}
	\begin{subfigure}{\textwidth}
		\centering
		\includegraphics{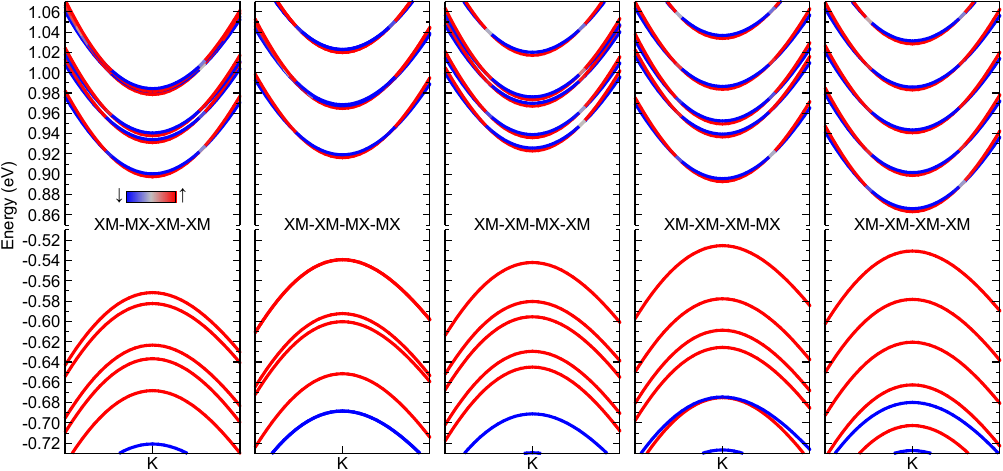}
	\end{subfigure}
	\caption{Band structure with a zoom around the K point.}
	\label{fig:bszoomK}
\end{figure}

\begin{figure}
	\centering    
	\includegraphics{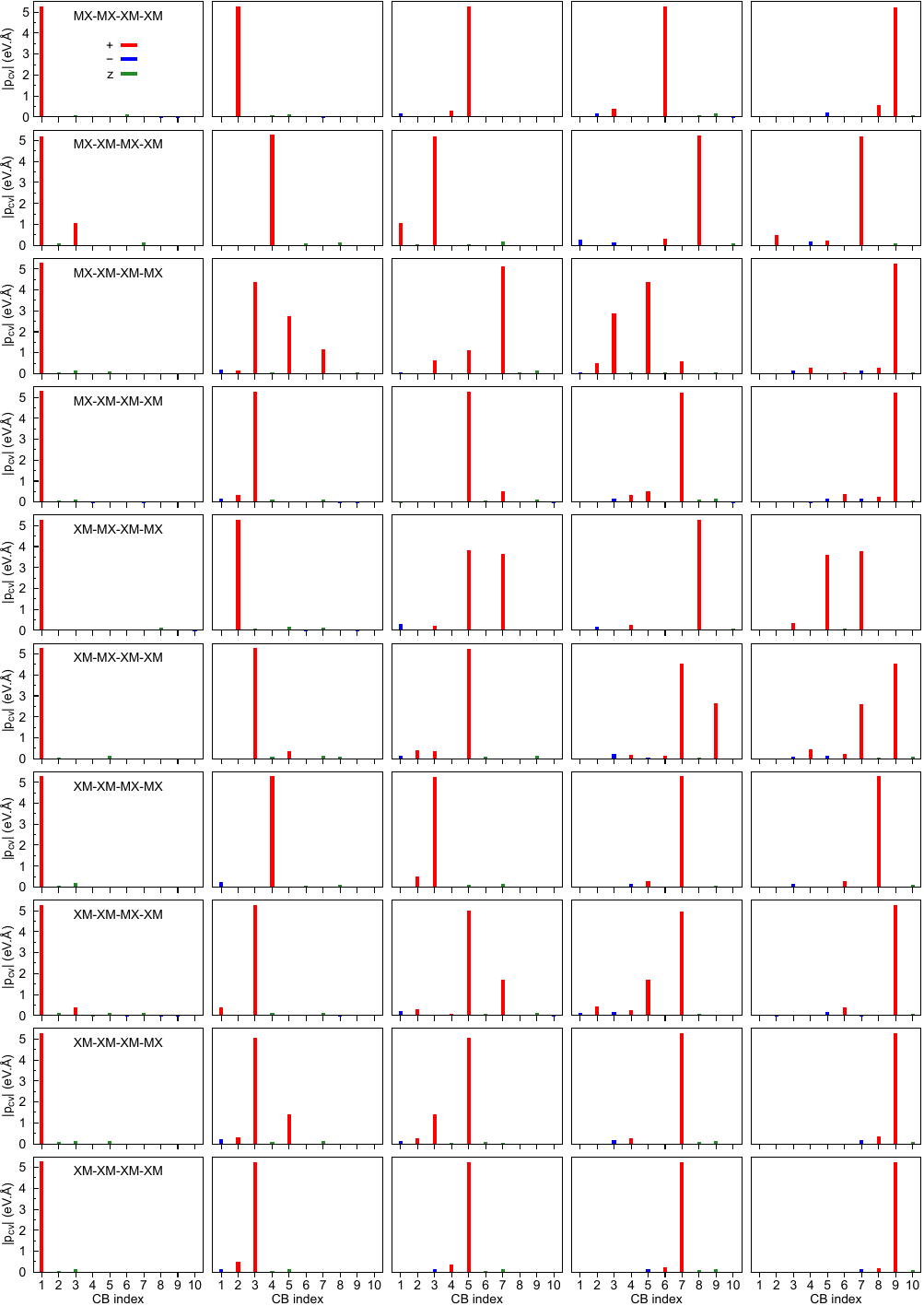}
	\caption{Absolute value of the dipole matrix elements between the valence bands with spin-up (different columns) to all conduction bands (labeled by index).}
	\label{fig:pcv_label}
\end{figure}

\begin{figure}
	\centering    \includegraphics{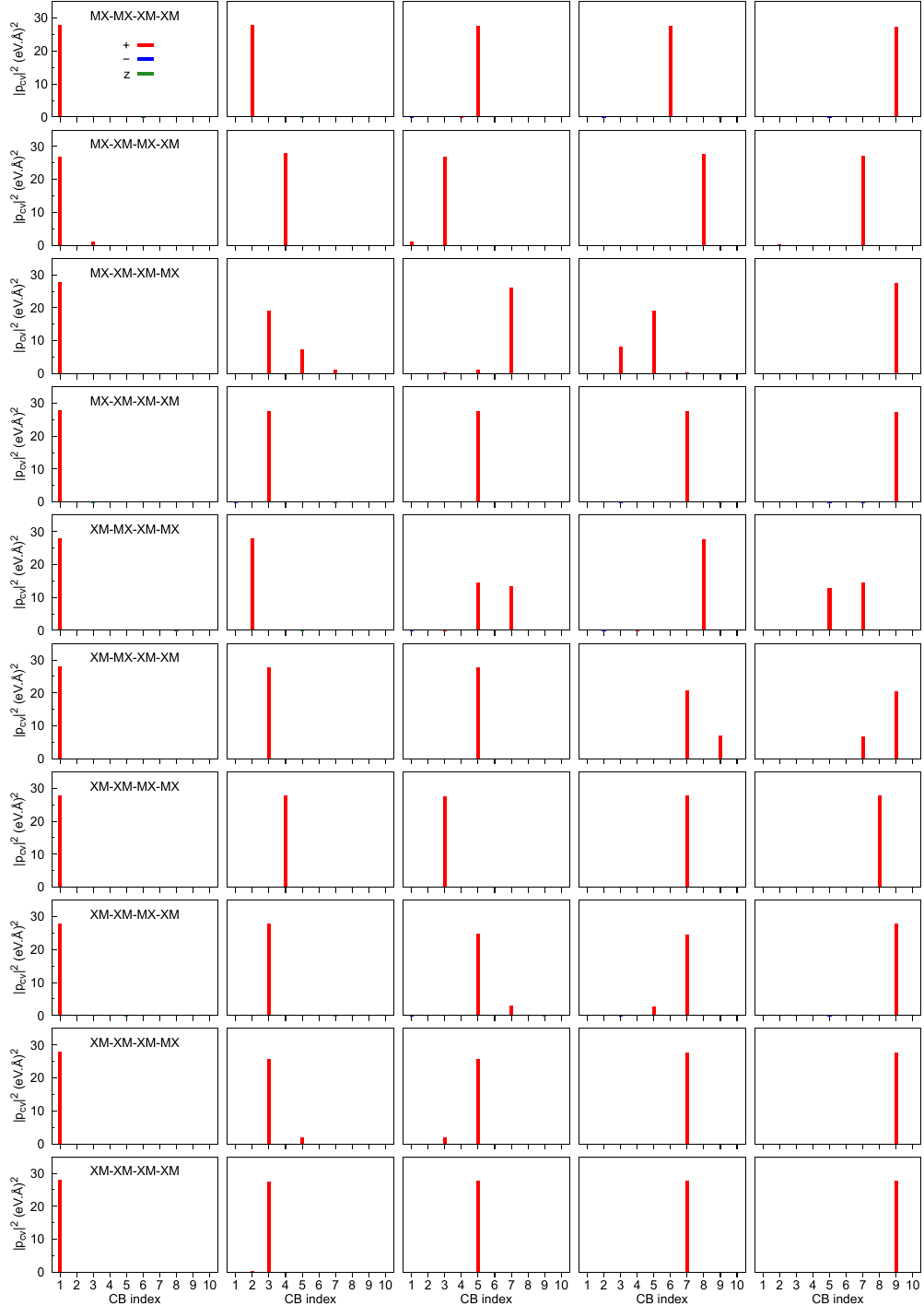}
	\caption{Same as Fig.~\ref{fig:pcv_label} but for the oscillator strength, i. e., $\left|p_{cv}\right|^2$.}
	\label{fig:pcv2_label}
\end{figure}

\begin{figure}
	\centering    \includegraphics{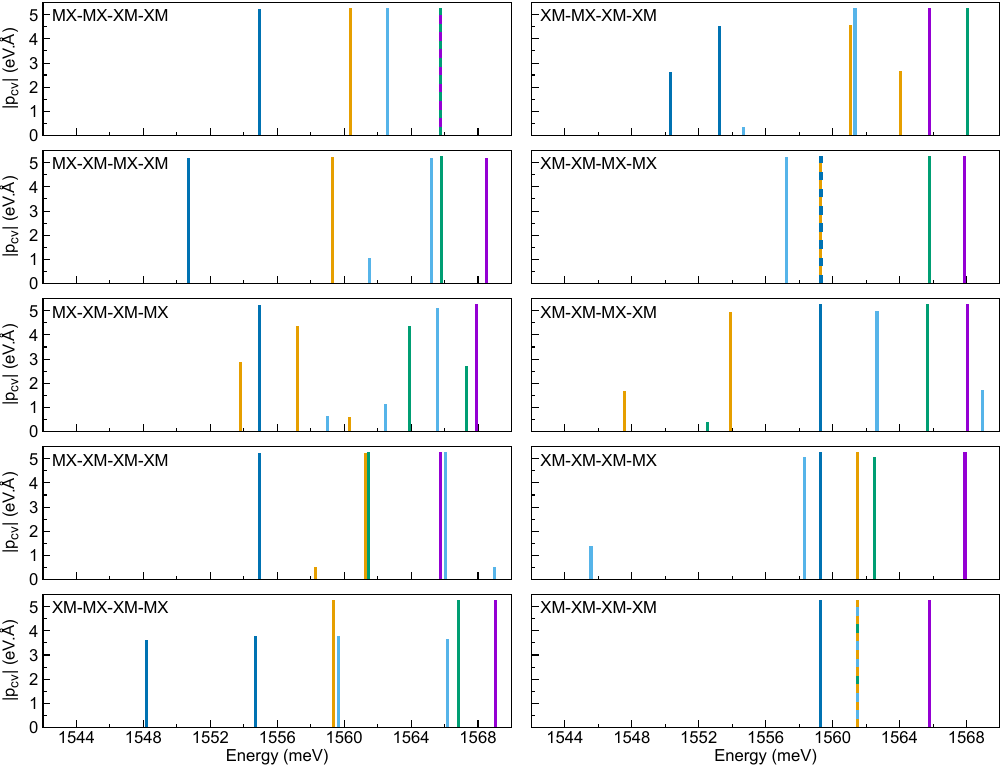}
	\caption{Absolute value of the dipole matrix elements for $s^+$ between the valence bands with spin-up and all conduction bands as a function of the transition energy.}
	\label{fig:pcv_energy}
\end{figure}

\begin{figure}
	\centering    \includegraphics{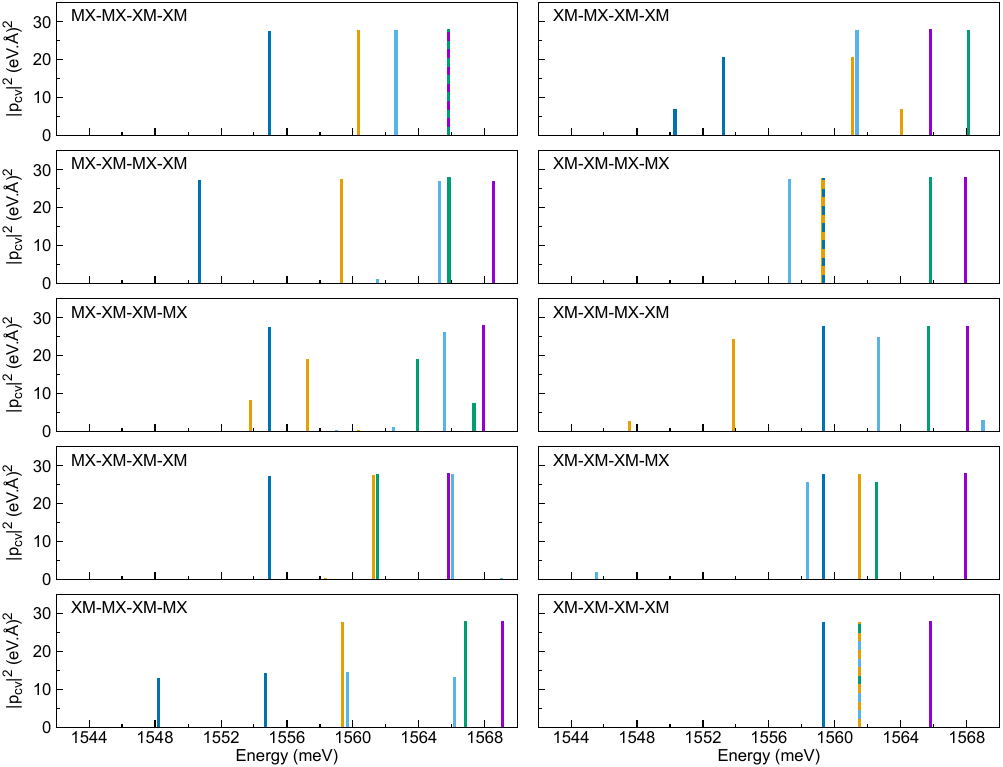}
	\caption{Same as Fig.~\ref{fig:pcv_energy} but for the oscillator strength, i. e., $\left|p_{cv}\right|^2$.}
	\label{fig:pcv2_energy}
\end{figure}

\begin{figure}
	\centering    
	\includegraphics{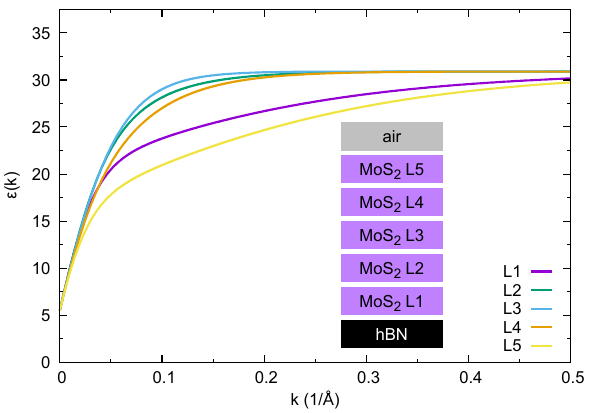}
	\caption{Layer dependent dielectric constants used in the exciton calculations.}
	\label{fig:eps}
\end{figure}

\begin{figure}
	\centering    
	\includegraphics{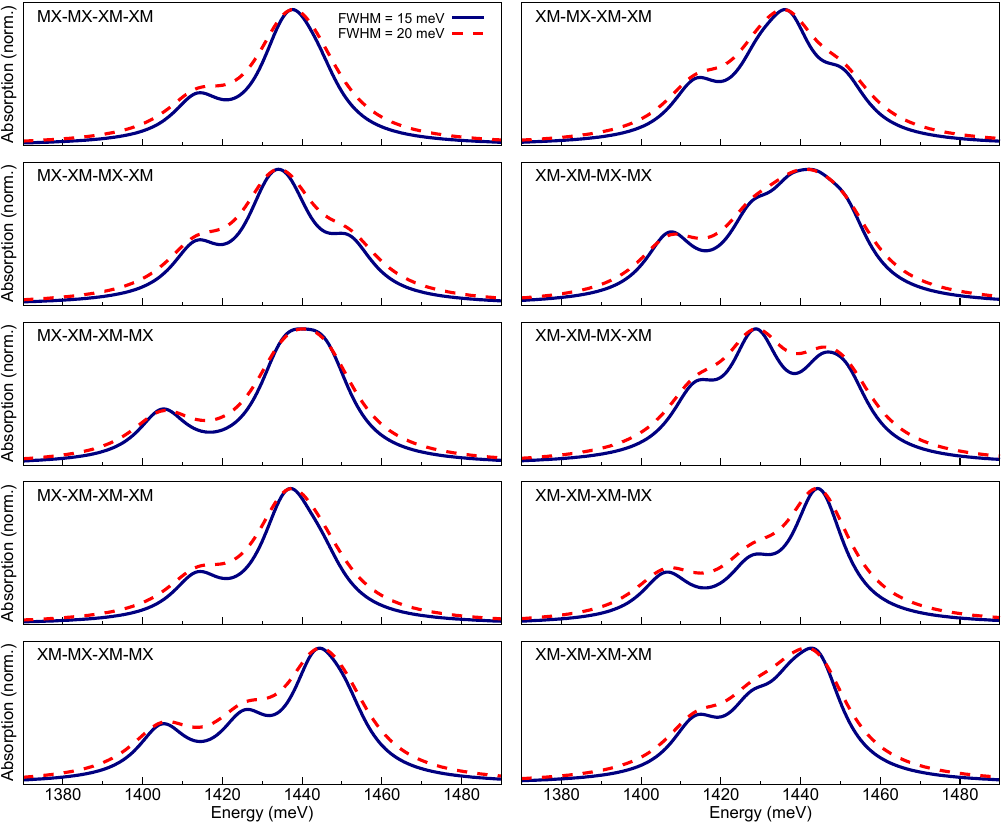}
	\caption{Exciton absorption spectra for the different stackings using a phenomenological Lorentzian broadening  with a full-width-half-maximum of 15 (20) meV shown by solid (dashed) lines.}
	\label{fig:abs}
\end{figure}

\begin{figure}
	\centering    
	\hspace*{1cm}\includegraphics[scale = 0.5]{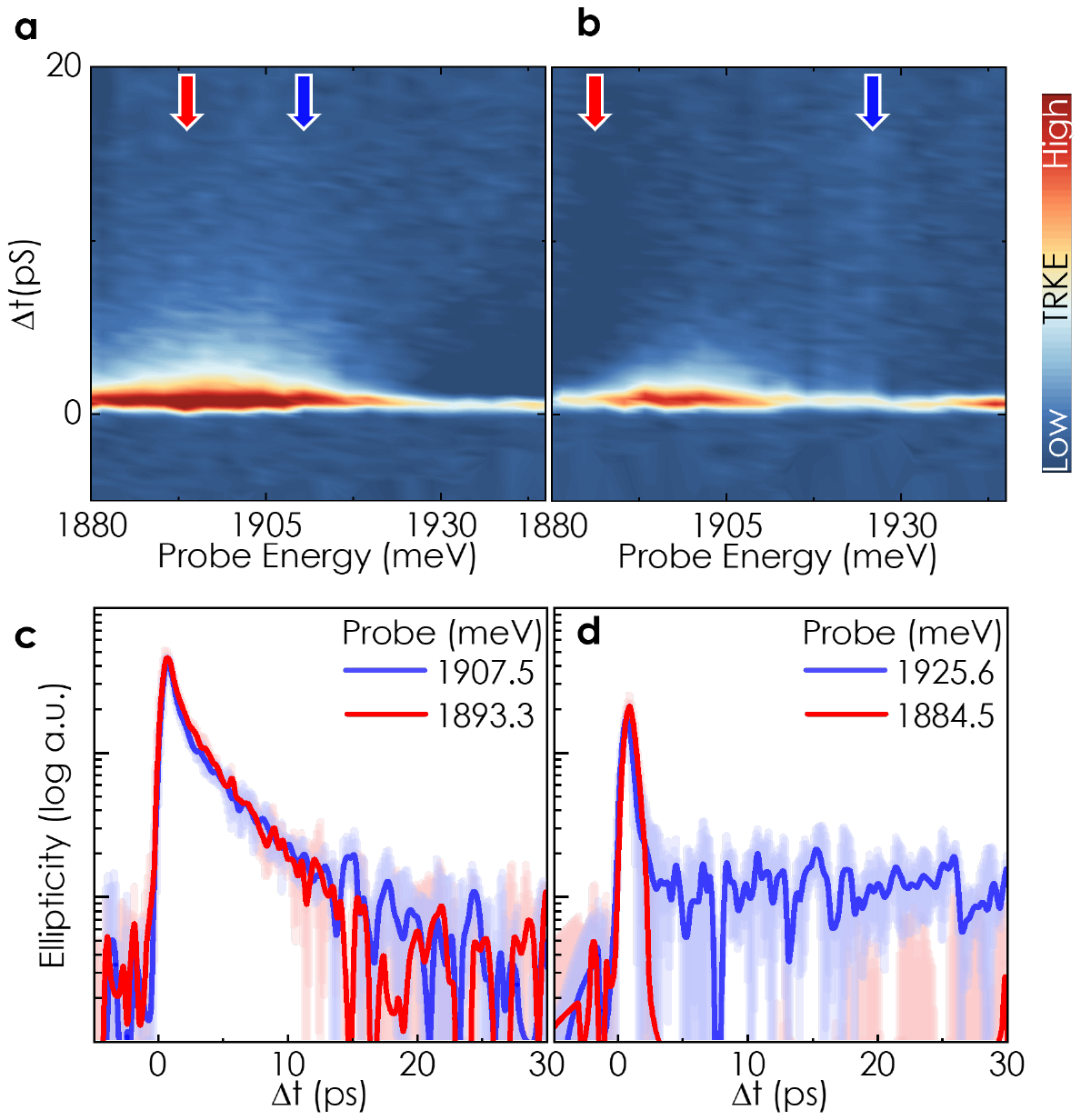}
	\caption{\textbf{a-b.} False color map of transient ellipticity as a function of probe energy from domain I and II, respectively. \textbf{c-d.} Transient reflection at selected energies, marked by colored arrows in the top panels (copied from Fig. 4 of the main text for completeness).}
	\label{fig:S27}
\end{figure}

\clearpage

\bibliographystyle{unsrt}
\bibliography{sn-bibliography}

\end{document}